\documentclass[nobibnotes,superscriptaddress,nofootinbib,twocolumn]{revtex4-1}

\usepackage{amsmath,amssymb}
\usepackage{graphicx}

\begin{document}
\title{Confronting predictive texture zeros in lepton mass matrices with current data}
\date{\today}

\author{Lu\'{\i}s M. Cebola}
\email{luismcebola@tecnico.ulisboa.pt}
\affiliation{Departamento de F\'{\i}sica and Centro de F\'{\i}sica
	Te\'{o}rica de Part\'{\i}culas - CFTP, Instituto Superior T\'{e}cnico,
	Universidade de Lisboa, Av. Rovisco Pais, 1049-001 Lisboa,
	Portugal}

\author{D.~Emmanuel-Costa}
\email{david.costa@tecnico.ulisboa.pt}
\affiliation{Departamento de F\'{\i}sica and Centro de F\'{\i}sica
	Te\'{o}rica de Part\'{\i}culas - CFTP, Instituto Superior T\'{e}cnico,
	Universidade de Lisboa, Av. Rovisco Pais, 1049-001 Lisboa,
	Portugal}

\author{R.~Gonz\'{a}lez Felipe}
\email{ricardo.felipe@tecnico.ulisboa.pt}
\affiliation{ISEL - Instituto Superior de Engenharia de Lisboa, Instituto Polit\'ecnico de Lisboa, Rua Conselheiro Em\'{\i}dio Navarro 1959-007 Lisboa,	Portugal}
\affiliation{Departamento de F\'{\i}sica and Centro de F\'{\i}sica
	Te\'{o}rica de Part\'{\i}culas - CFTP, Instituto Superior T\'{e}cnico,
	Universidade de Lisboa, Av. Rovisco Pais, 1049-001 Lisboa,
	Portugal}


\begin{abstract}
Several popular \emph{Ans\"{a}tze} of lepton mass matrices that contain texture zeros are confronted with current neutrino observational data. We perform a systematic $\chi^2$-analysis in a wide class of schemes, considering arbitrary Hermitian charged lepton mass matrices and symmetric mass matrices for Majorana neutrinos or Hermitian mass matrices for Dirac neutrinos. Our study reveals that several patterns are still consistent with all the observations at 68.27\% confidence level, while some others are disfavored or excluded by the experimental data. The well-known Frampton-Glashow-Marfatia two-zero textures, hybrid textures and parallel structures, among others, are considered.
\end{abstract}
\maketitle

\section{Introduction}
\label{sec:intro}

In the absence of a convincing theory to explain the origin of the lepton flavor structure, different approaches have been pursued to address this question. Among them, the imposition of texture zeros in the lepton mass matrices has been quite popular. The reason is two-fold. The vanishing of some matrix elements obviously reduces the number of free parameters, thus increasing, in some cases, the predictive power of the flavor patterns. Furthermore, texture zeros can naturally appear in theories with an extended scalar sector in the presence of Abelian symmetries~\cite{Grimus:2004hf,Felipe:2014vka}. Thus, the study of the phenomenological implications of lepton mass matrices with vanishing elements is well motivated on theoretical grounds.

During the last years, our knowledge of neutrino masses and leptonic mixing has been enriched thanks to the data accumulated from several solar, atmospheric, reactor and accelerator neutrino experiments~\cite{Capozzi:2013csa,Forero:2014bxa,Gonzalez-Garcia:2014bfa}, as well as to cosmological observations~\cite{Planck:2015xua}. Furthermore, an improved sensitivity to the Dirac CP phase has emerged from the complementarity of accelerator and reactor neutrino data. It is conceivable that leptonic CP violation is observed in current and next-generation neutrino oscillation experiments, which makes the search for such effects one of the main goals of the future research in neutrino physics~\cite{Branco:2011zb}.

It has been known for some time that, in the flavor basis where the charged lepton mass matrix is diagonal, neutrino mass matrices with more than two independent zero entries are not compatible with neutrino oscillation data, while seven patterns with two zeros are viable, as shown by Frampton, Glashow and Marfatia (FGM) in Ref.~\cite{Frampton:2002yf}. The latter contain four complex parameters, from which nine physical quantities should be determined (three neutrino masses, three mixing angles, one Dirac CP phase and two Majorana phases), assuming that light neutrinos are Majorana particles. More recently, the aforementioned two-zero textures have been scrutinized (see e.g. Refs.~\cite{Fritzsch:2011qv,Meloni:2012sx,Meloni:2014yea}). Other predictive textures can be envisaged as well in the flavor basis. The so-called hybrid textures~\cite{Kaneko:2005yz}, having one texture zero and two equal nonzero elements, contain the same number of physical parameters as the FGM textures. A systematic analysis of such hybrid textures has been presented in Ref.~\cite{Liu:2013oxa}, in which the authors concluded that 39 patterns for Majorana neutrinos are compatible with current neutrino oscillation data at the $3\sigma$ confidence level (C.L.).

Restrictive patterns for the lepton mass matrices can also be constructed when the charged lepton mass matrix is not diagonal. For instance, one can consider scenarios in which both matrices exhibit a ``parallel" structure~\cite{Branco:1999nb,Branco:2007nn} with the vanishing matrix elements located at the same positions~\cite{Nishiura:1999yt,Xing:2002sb} (see also Ref.~\cite{Gupta:2012dma} and references therein). Recently, a detailed survey of texture zeros in lepton mass matrices has been performed, for both Dirac and Majorana neutrinos, considering parallel and non-parallel matrix structures~\cite{Ludl:2014axa}. In the latter study, however, the Dirac phase was not included in the numerical $\chi^2$-analysis, which was carried out at the $5\sigma$ C.L..

In this work, we perform a detailed $\chi^2$-analysis of several popular \emph{Ans\"{a}tze} for lepton mass matrices that contain texture zeros. We aim at determining whether such patterns are consistent or not with current neutrino oscillation data at the $1\sigma\, (68.27\%)$ C.L.. In particular, the well-known FGM two-zero textures, the hybrid textures, as well as parallel structures will be analyzed. In our fitting procedure, we take into account six neutrino observables, namely, the two mass-squared differences, the three mixing angles, and the Dirac CP-violating phase. We also impose the recent cosmological bound on the sum of the neutrino masses~\cite{Planck:2015xua}. 

The paper is organized as follows. In Sec.~\ref{sec:chianalysis}, we briefly explain our strategy for the numerical analysis and minimization procedure. Then, we proceed in Sec.~\ref{sec:fgmtextures} to revisit the FGM two-zero textures for Majorana neutrinos. Two-zero textures for the lepton mass matrices in the case of Dirac neutrinos are also considered.  Section~\ref{sec:hybridtextures} is devoted to the systematic $\chi^2$-analysis of hybrid textures containing one-zero texture and two equal nonzero elements, for both Majorana and Dirac neutrinos. Parallel structures with two and three zeros are studied in Sec.~\ref{sec:partextures}. In Sec.~\ref{sec:nnitextures}, predictive neutrino textures in combination with a charged lepton mass matrix exhibiting the so-called nearest-neighbor-interaction (NNI) form are considered. Finally, our concluding remarks are given in Sec.~\ref{sec:summary}.

\section{Strategy for the numerical analysis}
\label{sec:chianalysis}

Leptonic mixing is described by the Pontecorvo, Maki, Nakagawa and Sakata (PMNS) matrix~\cite{pmns}, which, in the standard parametrization, can be written as~\cite{Agashe:2014kda}
\begin{widetext}
\begin{equation}\label{pmns-pdg}
\mathbf{U} = \left( 
\begin{array}{ccc}
c_{12}c_{13} & s_{12}c_{13} & s_{13}e^{-i\delta } \\ 
-s_{12}c_{23}-c_{12}s_{23}s_{13}e^{i\delta } & \quad
c_{12}c_{23}-s_{12}s_{23}s_{13}e^{i\delta }\quad & s_{23}c_{13} \\ 
s_{12}s_{23}-c_{12}c_{23}s_{13}e^{i\delta } & 
-c_{12}s_{23}-s_{12}c_{23}s_{13}e^{i\delta } & c_{23}c_{13}
\end{array}
\right) \cdot \mathrm{diag\, }(1,e^{i\alpha_{21}/2}, e^{i\alpha_{31}/2}),  
\end{equation}
\end{widetext}
where $c_{ij}\equiv \cos \theta _{ij}\ ,\ s_{ij}\equiv \sin \theta _{ij}\ $,
with all the angles $\theta _{ij}$ in the first quadrant,  $\delta$ 
is the Dirac CP phase, and $\alpha_{21}, \alpha_{31}$ are two Majorana phases. The unitary matrix $\mathbf{U}$ in Eq.~\eqref{pmns-pdg} relates the mass eigenstate neutrinos $\nu_i \ (i = 1,2,3)$ to the flavor eigenstate neutrinos $\nu_f \ (f=e,\mu,\tau)$. 

For Majorana neutrinos, the neutrino mass matrix $\mathbf{m}_\nu$ is a $3\times3$ complex symmetric matrix, which can be diagonalized by the unitary transformation $\mathbf{U}_{\nu L}^\dagger \mathbf{m}_\nu \mathbf{U}^{\ast}_{\nu L}=\mathrm{diag\,}(m_1,m_2,m_3)$, with $\mathbf{U}_{\nu L}$ a unitary matrix and the neutrino masses $m_i$ real and positive. If neutrinos are Dirac particles, then the corresponding unitary transformation is $\mathbf{U}_{\nu L}^\dagger \mathbf{m}_\nu \mathbf{U}_{\nu R}=\mathrm{diag\,}(m_1,m_2,m_3)$, in analogy to the charged leptons, for which the mass matrix $\mathbf{m}_\ell$ is diagonalized by  $\mathbf{U}_{\ell L}^\dagger \mathbf{m}_\ell \mathbf{U}_{\ell R}=\mathrm{diag\,}(m_e,m_\mu,m_\tau)$. The leptonic mixing matrix $\mathbf{U}$ is then given by $\mathbf{U}=\mathbf{U}_{\ell L}^\dagger \mathbf{U}_{\nu L}$, which can always be parametrized in the form of Eq.~\eqref{pmns-pdg}.

The absolute scale of neutrino masses is not yet known and there are two possible orderings of the light neutrino masses: normal ordering (NO) with $m_1<m_2<m_3$ or inverted ordering (IO) with $m_3<m_1<m_2$. The spectrum may vary from hierarchical to quasi-degenerate masses.  Nevertheless, cosmological observations place a stringent upper bound on the sum of the masses. Assuming three species of degenerate massive neutrinos and a $\Lambda$CDM model, the Planck collaboration has released the bound~\cite{Planck:2015xua}  
\begin{equation}
\label{planck}
\sum_i m_i < 0.23~\text{eV}\quad \text{(95\% C.L.)},
\end{equation}
obtained from a combined analysis of data.\footnote{Similar bounds are inferred from other cosmological observations. For instance, the median value $\sum_i m_i=0.32 \pm 0.11$~eV has been obtained by the South Pole Telescope collaboration, with a $3\sigma$ detection of a positive sum and $\sum_i m_i \in [0.01,0.63]$~eV at 99.7\% C.L.~\cite{Hou:2012xq}.} Although this bound is not definite and requires confirmation by forthcoming experiments, its inclusion in the analysis of neutrino mass models may lead to important conclusions about the viability of a given model.

In this work, we shall perform a $\chi^2$-analysis using the standard $\chi^2$-function
\begin{equation}\label{chisquared}
\chi^2(x) = \sum_{i}
\frac{(\mathcal{P}_i(x)-\overline{\mathcal{O}}_i)^2}{\sigma_i^2},
\end{equation}
where $x$ denotes the physical input parameters (in our case, the matrix elements of the lepton mass matrices), $\mathcal{P}_i(x)$ are the
predictions of the \emph{Ans\"{a}tze} for the observables $\mathcal{O}_i$, $\overline{\mathcal{O}}_i$ are the best-fit values of $\mathcal{O}_i$, and $\sigma_i$ are their corresponding $1\sigma$ errors. In our study, we make use of the current neutrino  parameters at $1\sigma$, obtained in Ref.~\cite{Forero:2014bxa} from the global fit of neutrino oscillation data. Furthermore, we impose the cosmological constraint on the sum of the neutrino masses given in Eq.~\eqref{planck}.

We shall fit the zero-textures of lepton mass matrices taking into account six observables: the mass-squared differences $\Delta m_{21}^2,\, \Delta m_{31}^2$, the mixing angles $\mathrm{sin}^2\theta_{12},\, \mathrm{sin}^2\theta_{23},\,
\mathrm{sin}^2\theta_{13}$, and the Dirac CP phase $\delta$. Since  the Majorana phases are presently not constrained, we do not include them in the analysis. A given texture is considered to agree well with the experimental data if the model predictions for the physical observables in Eq.~\eqref{chisquared} are within the $1\sigma$ interval given in Table~\ref{tab:nudata}. Thus, $\chi^2_{min}\lesssim 6$ is a necessary condition for a pattern to be consistent with all observations. 

\begin{table}[ht]\centering
	\begin{tabular}{lc}
		\hline
		Parameter & Best fit $\pm$ $1\sigma$\\
		\hline
		$\Delta m^2_{21}\: [10^{-5}\,\text{eV}^2]$
		& 7.60$^{+0.19}_{-0.18}$ \\[2mm] 
		$|\Delta m^2_{31}|\: [10^{-3}\,\text{eV}^2]$ (NO)
		&  2.48$^{+0.05}_{-0.07}$ \\
		\phantom{$|\Delta m^2_{31}|\: [10^{-3}\text{eV}^2]$ } (IO)
		&  2.38$^{+0.05}_{-0.06}$  \\[2mm] 
		$\sin^2\theta_{12} / 10^{-1}$
		& 3.23$\pm$0.16 \\[2mm]  
		$\sin^2\theta_{23} / 10^{-1}$ (NO)
		& 5.67$^{+0.32}_{-1.24}$  \\
		\phantom{$\sin^2\theta_{23} / 10^{-1}$ } (IO)
		& 5.73$^{+0.25}_{-0.39}$ \\[2mm]
		$\sin^2\theta_{13} / 10^{-2}$ (NO)
		& 2.26$\pm$0.12  \\
		\phantom{$\sin^2\theta_{13} / 10^{-2}$ } (IO)
		& 2.29$\pm$0.12  \\[2mm]
		$\delta/\pi$\quad (NO)
		& 1.41$^{+0.55}_{-0.40}$ \\
		\phantom{$\delta/\pi$ }\quad (IO)	
		& 1.48$\pm$0.31 \\	
		\hline
	\end{tabular}
	\caption{\label{tab:nudata} Neutrino oscillation parameters at $68.27\%$ C.L. taken from Ref.~\cite{Forero:2014bxa}. The upper and lower rows in $\Delta m^2_{31}$, $\sin^2\theta_{23}$, $\sin^2\theta_{13}$,and $\delta$  correspond to normal (NO) and inverted (IO) neutrino mass ordering, respectively.}
\end{table}

We remark that our approach to the determination of the charged lepton masses slightly differs from that of Ref.~\cite{Ludl:2014axa}. In our  search for viable charged-lepton mass matrices, we always require that the eigenvalues of the input mass matrix correctly reproduce the central values of the charged lepton masses~\cite{Agashe:2014kda}, i.e.
\begin{align}
m_e&=0.510998928~\text{MeV},\;\nonumber\\
m_\mu&=105.6583715~\text{MeV},\;\\
m_\tau&=1776.82~\text{MeV}.\nonumber
\end{align}
The minimization of the $\chi^2$-function is carried out with respect to the 6 neutrino observables using the MINUIT package~\cite{James:1975dr,root}. To improve the quality of the minima, this procedure is repeated $10^4$ times, with randomly chosen initial charged lepton and neutrino mass matrices. Clearly, in the weak basis where the charged lepton mass matrix is diagonal and real (flavor basis), one has $\mathbf{m}_\ell=\mathrm{diag\,}(m_e,m_\mu,m_\tau)$ and thus this matrix is fixed. Moreover, one can easily show that the absolute value of any matrix element of $\mathbf{m}_\nu$ is always smaller than the largest neutrino mass, i.e.
$|(\mathbf{m}_\nu)_{ij}|<\max_k\, (m_k)$. Therefore, the cosmological bound in Eq.~\eqref{planck} implies $|(\mathbf{m}_\nu)_{ij}|\lesssim 0.08$~eV.
	
\section{FGM textures}
\label{sec:fgmtextures}

In this section we revisit the well-known FGM patterns for lepton mass matrices~\cite{Frampton:2002yf}, consisting of $3\times3$ Majorana neutrino mass matrices $\mathbf{m}_\nu$ with two zero elements in the charged lepton flavor basis with $\mathbf{m}_\ell=\text{diag}\,(m_e,m_\mu,m_\tau)$. We shall also consider the case of Dirac neutrinos, for which the matrix $\mathbf{m}_\nu$ is Hermitian.

For Majorana neutrinos, the mass matrix $\mathbf{m}_\nu$ is a symmetric matrix with six independent complex entries. There are $6!/[n! (6 - n)!]$ different textures, each containing $n$ independent texture zeros. One can show that any pattern of $\mathbf{m}_\nu$ with more than two independent zeros ($n> 2$) is not compatible with current neutrino oscillation data. For $n=2$,
there are fifteen two-zero textures of $\mathbf{m}_\nu$, which can be classified into
six categories ($\mathbf{A}, \mathbf{B}, \mathbf{C}, \mathbf{D}, \mathbf{E},
\mathbf{F}$), according to their physical predictions:

\begin{align}
\label{FGMtextures}
\mathbf{A}_1:
\begin{pmatrix} 0 & 0 & \ast \\ 0 & \ast & \ast \\
\ast & \ast &
\ast
\end{pmatrix}
\,,\quad
\mathbf{A}_2:
\begin{pmatrix}
0 & \ast & 0 \\ \ast & \ast & \ast \\ 0 & \ast & \ast
\end{pmatrix}
\,,\quad
\mathbf{B}_1:
\begin{pmatrix}
\ast & \ast & 0 \\ \ast & 0 & \ast \\ 0 & \ast & \ast
\end{pmatrix}\,,\nonumber\\[3mm]
\mathbf{B}_2:
\begin{pmatrix}
\ast & 0 & \ast \\ 0 & \ast & \ast \\ \ast & \ast & 0
\end{pmatrix}\,,\quad
\mathbf{B}_3:
\begin{pmatrix}
\ast & 0 & \ast \\ 0 & 0 & \ast \\ \ast & \ast & \ast
\end{pmatrix}\,,\quad
\mathbf{B}_4:
\begin{pmatrix}
\ast & \ast & 0 \\ \ast & \ast & \ast \\ 0 & \ast & 0
\end{pmatrix}\,,\nonumber\\[3mm]
\mathbf{C}:
\begin{pmatrix}
\ast & \ast & \ast \\ \ast & 0 & \ast \\ \ast & \ast & 0
\end{pmatrix}\,,\quad
\mathbf{D}_1:
\begin{pmatrix}
\ast & \ast & \ast \\ \ast & 0 & 0 \\ \ast & 0 & \ast
\end{pmatrix}\,,\quad
\mathbf{D}_2:
\begin{pmatrix}
\ast & \ast & \ast \\ \ast & \ast & 0 \\ \ast & 0 & 0
\end{pmatrix}\,,\nonumber\\[3mm]
\mathbf{E}_1:
\begin{pmatrix}
0 & \ast & \ast \\ \ast & 0 & \ast \\ \ast & \ast & \ast
\end{pmatrix}\,,\quad
\mathbf{E}_2:
\begin{pmatrix}
0 & \ast & \ast \\ \ast & \ast & \ast \\ \ast & \ast & 0
\end{pmatrix}\,,\quad
\mathbf{E}_3:
\begin{pmatrix}
0 & \ast & \ast \\ \ast & \ast & 0 \\ \ast & 0 & \ast
\end{pmatrix}\,,\nonumber\\[3mm]
\mathbf{F}_1:
\begin{pmatrix}
\ast & 0 & 0 \\ 0 & \ast & \ast \\ 0 & \ast & \ast
\end{pmatrix} \,,\quad
\mathbf{F}_2:
\begin{pmatrix}
\ast & 0 & \ast \\ 0 & \ast & 0 \\ \ast & 0 & \ast
\end{pmatrix}\,,\quad
\mathbf{F}_3:
\begin{pmatrix}
\ast & \ast & 0 \\ \ast & \ast & 0 \\ 0 & 0 & \ast
\end{pmatrix}\,.\nonumber\\
\end{align}
Here, the symbol ``$\ast$" stands for arbitrary nonzero matrix elements. Clearly, the  matrices $\mathbf{F}_i$ can be straightforwardly excluded since they lead to the decoupling of one generation and thus are not experimentally viable.

\begin{table*}[ht]
\begin{tabular}{clccccccc}
	\hline
Majorana $\mathbf{m}_\nu$	&\;\;$\chi^2_{min}$ NO &\;\;\;\; (IO)\;\;\;\;  &\; $\Delta m^2_{21}$\;  &\; $\Delta m^2_{31}$\;  & \;\;\;$\theta_{12}$\;\;\; &\;\;\;$\theta_{23}$\;\;\;  &\;\;\;$\theta_{13}$\;\;\;  &\; \;\;$\delta$\;\;\; \\ \hline	
$\mathbf{A}_1$	&  $2.92\times10^{-1}$ & ($3.81\times10^{2}$) & $\checkmark$ &$\checkmark$  & $\checkmark (\times)$ & $\checkmark (\times)$ & $\checkmark$ & $\checkmark (\times)$ \\ 
$\mathbf{A}_2$	& $1.23\times10^{-2}$ & ($3.14\times10^{2}$) & $\checkmark$ &$\checkmark$  & $\checkmark (\times)$ & $\checkmark (\times)$ & $\checkmark$ & $\checkmark (\times)$\\ 
$\mathbf{B}_1$	&  $8.39\times10^{-1}$ & ($4.04\times10^{-3}$) & $\checkmark$ &  $\checkmark$ &  $\checkmark$ &  $\checkmark$ &  $\checkmark$ &  $\checkmark$ \\ 
$\mathbf{B}_2$	& $3.39\times10^{-2}$ &($1.02\times10^{1}$) & $\checkmark$ &  $\checkmark$ &  $\checkmark$ &  $\checkmark (\times)$ &  $\checkmark$ &  $\checkmark$  \\ 
$\mathbf{B}_3$	& $9.12\times10^{-1}$ & ($3.45\times10^{-3}$)  & $\checkmark$ &  $\checkmark$ &  $\checkmark$ &  $\checkmark$ &  $\checkmark$ &  $\checkmark$ \\ 
$\mathbf{B}_4$	& $2.10\times10^{-2}$ & ($1.11\times10^{1}$) & $\checkmark$ &  $\checkmark$ &  $\checkmark$ &  $\checkmark (\times)$ &  $\checkmark$ &  $\checkmark$ \\ 
$\mathbf{C}$	& $6.20\times10^{2}$ & ($1.04\times10^{-1}$) & $\checkmark$ &  $\times (\checkmark)$ &  $\checkmark$ &  $\checkmark$ &  $\times (\checkmark)$ &  $ \checkmark$  \\ 
$\mathbf{D}_1$	& $1.33\times10^{2}$ & ($3.43\times10^{1}$) & $\checkmark$ &  $\checkmark$ &  $\times$ &  $\times$ &  $\checkmark$ &  $\checkmark (\times)$  \\ 
$\mathbf{D}_2$	& $2.82\times10^{2}$ & ($4.88\times10^{1}$) & $\checkmark$ &  $\checkmark$ &  $\times$ &  $\times$ &  $\checkmark$ &  $\checkmark (\times)$ \\ 
$\mathbf{E}_1$	& $1.40\times10^{1}$ & ($1.15\times10^{2}$) & $\checkmark$ &  $\checkmark$ &  $\checkmark (\times)$ &  $\times (\checkmark)$ &  $\checkmark$ &  $\checkmark$  \\ 
$\mathbf{E}_2$	& $1.03\times10^{2}$ & ($1.14\times10^{2}$) &$\checkmark$ &  $\checkmark$ &  $\checkmark (\times)$ &  $\times (\checkmark)$ &  $\checkmark$ &  $\checkmark$  \\ 
$\mathbf{E}_3$	& $2.09\times10^{1}$ & ($1.17\times10^{2}$) & $\checkmark$ &  $\checkmark$ &  $\checkmark (\times)$ &  $\times (\checkmark)$ &  $\checkmark$ &  $\checkmark$ \\
\hline
\end{tabular} 
\caption{\label{tab:FGM}The minimum of $\chi^2$ for the FGM zero-textures of the neutrino mass matrix with a normal (inverted) mass ordering. We use the data given in Table~\ref{tab:nudata} and impose the upper bound on the sum of neutrino masses of Eq.~\eqref{planck}. In all cases, the charged lepton mass matrix is $\mathbf{m}_\ell=\text{diag}\,(m_e,m_\mu,m_\tau)$. We also indicate with a check mark or a cross whether the predictions are or not within the $1\sigma$ interval given in Table~\ref{tab:nudata}.}
\end{table*}

\begin{table*}[ht]
	\begin{tabular}{clccccccc}
		\hline
		Dirac $\mathbf{m}_\nu$	&\;\;$\chi^2_{min}$ NO &\;\;\;\; (IO)\;\;\;\;  &\; $\Delta m^2_{21}$\;  &\; $\Delta m^2_{31}$\;  & \;\;\;$\theta_{12}$\;\;\; &\;\;\;$\theta_{23}$\;\;\;  &\;\;\;$\theta_{13}$\;\;\;  &\; \;\;$\delta$\;\;\; \\ \hline	
		$\mathbf{C}$	& $6.19\times10^{2}$ & ($1.04\times10^{-1}$) & $\checkmark$ &  $\times (\checkmark)$ &  $\checkmark$ &  $\checkmark$ &  $\times (\checkmark)$ &  $ \checkmark$ \\ 
		$\mathbf{E}_1$	& $1.40\times10^{1}$ & ($1.15\times10^{2}$) & $\checkmark$ &  $\checkmark$ &  $\checkmark (\times)$ &  $\times (\checkmark)$ &  $\checkmark$ &  $\checkmark(\times)$ \\ 
		$\mathbf{E}_2$	& $1.03\times10^{2}$ & ($1.14\times10^{2}$) &$\checkmark$ &  $\checkmark$ &  $\times$ &  $\times (\checkmark)$ &  $\checkmark$ &  $\checkmark(\times)$ \\ 
		\hline
	\end{tabular} 
	\caption{\label{tab:FGM-Dirac}As in Table~\ref{tab:FGM}, for the case of Dirac neutrinos. We present only the patterns for which the Dirac phase $\delta$ is different from 0 or $\pi$, leading to leptonic CP violation.}
\end{table*}

Our results are presented in Table~\ref{tab:FGM}, in which the minimum of $\chi^2$ for each FGM texture  with a normal or inverted neutrino mass ordering is given. The results are obtained using the current neutrino oscillation data of Table~\ref{tab:nudata} and imposing the upper bound on the sum of neutrino masses given in Eq.~\eqref{planck}. We indicate with a check mark or a cross whether the texture predictions are or not within the $1\sigma$ interval given in Table~\ref{tab:nudata}. Note that, in order to ease the reading of the table, whenever a given observable is simultaneously compatible (or incompatible) with data for NO and IO, we just indicate it with a single symbol, i.e. with a check mark (or a cross). Henceforth, this notation will be used in all tables.

From Table~\ref{tab:FGM} we conclude that patterns $\mathbf{A}_{1,2}$ and $\mathbf{B}_{1,2,3,4}$ are allowed for NO, while only patterns $\mathbf{B}_{1,3}$ and $\mathbf{C}$ are compatible with neutrino oscillation data for an IO mass spectrum at the $1\sigma$ level.\footnote{The seven matrices were previously found to be compatible with neutrino oscillation data at the $1\sigma$ level for NO and IO mass spectrum~\cite{Fritzsch:2011qv}.} We remark that, if the stringent upper bound on the sum of neutrino masses given in Eq.~\eqref{planck} is relaxed, pattern  $\mathbf{C}$ is also allowed for a NO neutrino mass spectrum~\cite{Meloni:2014yea}. In the latter case, we obtain $\chi^2_{min} \simeq 0.32$ with $\sum_i m_i < 1$~eV.

For completeness, in Figs.~\ref{fig:fig-A1}-\ref{fig:fig-C} of Appendix~\ref{appendix1}, we present the probability distribution of the six neutrino observables, obtained for the seven viable FGM textures  $\mathbf{A}_{1,2}$, $\mathbf{B}_{1,2,3,4}$ and $\mathbf{C}$, for both NO and IO mass spectra. We notice that textures in the same category lead in general to similar physical predictions for the observables.

We now consider the case of Dirac neutrinos. We analyze again the two-zero textures given in Eq.~\eqref{FGMtextures}. These patterns have been recently studied for Dirac neutrinos in Ref.~\cite{Liu:2012axa}, where the authors concluded that only the patterns $\mathbf{A}_{1,2}$ and $\mathbf{C}$ are compatible with the oscillation data at the $2\sigma$ level. 

First we note that by redefining the right-handed neutrino fields we can assume, without loss of generality, that the mass matrix $\mathbf{m}_\nu$ is Hermitian. Furthermore, it is straightforward to show that if one off-diagonal matrix element is zero, then the invariant quantity $\mathcal{J}_{CP}=\text{Im}\, [\mathbf{U}_{12}\mathbf{U}_{23}\mathbf{U}_{13}^\ast \mathbf{U}_{22}^\ast]$ vanishes, and thus CP is conserved in the lepton sector. Therefore, only patterns $\mathbf{C}$, $\mathbf{E}_1$, and $\mathbf{E}_2$ can lead to leptonic CP violation, while $\delta=0$ or $\pi$ for the remaining two-zero patterns. 

In view of the above, we shall only present the results for patterns $\mathbf{C}$, $\mathbf{E}_1$, and $\mathbf{E}_2$. The minimum of $\chi^2$ is given in Table~\ref{tab:FGM-Dirac}. As can be seen from the table, there is essentially no difference with respect to the results obtained for Majorana neutrinos. Only pattern $\mathbf{C}$ with an inverted hierarchy is allowed by current data. Relaxing the cosmological bound on the sum of the neutrino masses, we conclude that a normal hierarchical neutrino spectrum is also allowed for pattern $\mathbf{C}$,  with   $\chi^2_{min} \simeq 0.29$ for $\sum_i m_i < 1$~eV. Notice also that the parameter counting for Hermitian Dirac matrices differs from that of symmetric Majorana matrices, since in the former case the counting depends on the position of the zeros. For two vanishing diagonal matrix entries, the matrix $\mathbf{m}_\nu$ contains at most seven real parameters. 

\section{Hybrid textures}
\label{sec:hybridtextures}

\begin{table*}[ht]
	\begin{tabular}{clccccccc}
		\hline
		Majorana $\mathbf{m}_\nu$ 	&\;\;$\chi^2_{min}$ NO &\;\;\;\; (IO)\;\;\;\;  &\; $\Delta m^2_{21}$\;  &\; $\Delta m^2_{31}$\;  & \;\;\;$\theta_{12}$\;\;\; &\;\;\;$\theta_{23}$\;\;\;  &\;\;\;$\theta_{13}$\;\;\;  &\; \;\;$\delta$\;\;\; \\ \hline	
		$\widehat{\mathbf{A}}_{1(13)}$	&  $1.78\times10^{-8}$ & ($1.40\times10^{2}$) & $\checkmark$ &$\checkmark$  & $\checkmark$ & $\checkmark (\times)$ & $\checkmark (\times)$ & $\checkmark$ \\ 
		$\widehat{\mathbf{A}}_{1(22)}$	& $1.65$ & ($7.16\times10^{-7}$) & $\checkmark$ &$\checkmark$  & $\checkmark$ & $\times(\checkmark)$ & $\checkmark$ & $\checkmark$\\ 
		$\widehat{\mathbf{A}}_{1(23)}$	&  $1.76\times10^{1}$ & ($6.44\times10^{-9}$) & $\checkmark$ &  $\checkmark$ &  $\checkmark$ &  $\times (\checkmark)$ &  $\checkmark$ &  $\checkmark$ \\ 
		$\widehat{\mathbf{A}}_{1(33)}$	& $9.11$ &($7.22\times10^{-6}$) & $\checkmark$ &  $\checkmark$ &  $\checkmark$ &  $\times ($\checkmark$)$ &  $\checkmark$ &  $\checkmark$  \\ 
		$\widehat{\mathbf{A}}_{2(12)}$	& $1.29\times10^{-8}$ & ($1.55\times10^{2}$)  & $\checkmark$ &  $\checkmark$ &  $\checkmark$ &  $\checkmark(\times)$ &  $\checkmark(\times)$ &  $\checkmark$ \\ 
		$\widehat{\mathbf{A}}_{2(22)}$	& $3.45$ & ($1.99\times10^{-6}$) & $\checkmark$ &  $\checkmark$ &  $\checkmark$ &  $\times (\checkmark)$ &  $\checkmark$ &  $\checkmark$ \\ 
		$\widehat{\mathbf{A}}_{2(23)}$	& $2.06\times10^{1}$ & ($4.75\times10^{-11}$) & $\checkmark$ &  $\checkmark$ &  $\checkmark$ &  $\times (\checkmark)$ &  $\checkmark$ &  $\checkmark$  \\ 
		$\widehat{\mathbf{A}}_{2(33)}$	& $7.34\times10^{-1}$ & ($9.89\times10^{-6}$) & $\checkmark$ &  $\checkmark$ &  $\checkmark$ &  $\checkmark$ &  $\checkmark$ &  $\checkmark$  \\ 
		\hline
	\end{tabular} 
	\caption{\label{tab:hybridA}The minimum of $\chi^2$ for the $\widehat{\mathbf{A}}$-type hybrid textures.}
\end{table*}

\begin{table*}[ht]
	\begin{tabular}{clccccccc}
		\hline
		Majorana $\mathbf{m}_\nu$ 	&\;\;$\chi^2_{min}$ NO &\;\;\;\; (IO)\;\;\;\;  &\; $\Delta m^2_{21}$\;  &\; $\Delta m^2_{31}$\;  & \;\;\;$\theta_{12}$\;\;\; &\;\;\;$\theta_{23}$\;\;\;  &\;\;\;$\theta_{13}$\;\;\;  &\; \;\;$\delta$\;\;\; \\ \hline	
		$\widehat{\mathbf{B}}_{1(11)}$	&  $6.63$ & ($3.14\times10^{2}$) & $\checkmark$ &$\checkmark$  & $\checkmark(\times)$ & $ \times$ & $\checkmark$ & $\checkmark(\times)$ \\ 
		$\widehat{\mathbf{B}}_{1(12)}$	&  $6.34\times10^{-1}$ & ($3.67\times10^{-7}$) & $\checkmark$ &$\checkmark$  & $\checkmark$ & $ \checkmark$ & $\checkmark$ & $\checkmark$ \\ 
		$\widehat{\mathbf{B}}_{1(23)}$	& $2.08\times10^{1}$ & ($1.31\times10^{-8}$) & $\checkmark$ &$\checkmark$  & $\checkmark$ & $\times (\checkmark)$ & $\checkmark$ & $\checkmark$\\ 
		$\widehat{\mathbf{B}}_{1(33)}$	&  $3.42\times10^{-5}$ & ($6.83$) & $\checkmark$ &  $\checkmark$ &  $\checkmark$ &  $\checkmark(\times)$ &  $\checkmark$ &  $\checkmark$ \\ 
		$\widehat{\mathbf{B}}_{2(11)}$	& $3.04\times10^{1}$ &($3.80\times10^{2}$) & $\checkmark$ &  $\checkmark$ &  $\checkmark(\times)$ &  $\times$ &  $\checkmark$ &  $\checkmark(\times)$  \\ 
		$\widehat{\mathbf{B}}_{2(13)}$	& $9.64\times10^{-3}$ &($7.34$) & $\checkmark$ &  $\checkmark$ &  $\checkmark$ &  $\checkmark (\times)$ &  $\checkmark$ &  $\checkmark$  \\ 
		$\widehat{\mathbf{B}}_{2(22)}$	& $7.18\times10^{-1}$ & ($1.72\times10^{-6}$)  & $\checkmark$ &  $\checkmark$ &  $\checkmark$ &  $\checkmark$ &  $\checkmark$ &  $\checkmark$ \\ 
		$\widehat{\mathbf{B}}_{2(23)}$	& $1.80\times10^2$ & ($9.61\times10^{-10}$) & $\checkmark$ &  $\checkmark$ &  $\checkmark$ &  $\times (\checkmark)$ &  $\checkmark$ &  $\checkmark$ \\ 
		$\widehat{\mathbf{B}}_{3(11)}$	& $5.52$ & ($1.08\times10^{2}$) & $\checkmark$ &  $\checkmark$ &  $\checkmark(\times)$ &  $\times(\checkmark)$ &  $\checkmark$ &  $\checkmark$  \\ 
		$\widehat{\mathbf{B}}_{3(13)}$	& $5.59\times10^{-1}$ & ($2.69\times10^{-5}$) & $\checkmark$ &  $\checkmark$ &  $\checkmark$ &  $\checkmark$ &  $\checkmark$ &  $\checkmark$  \\ 
		$\widehat{\mathbf{B}}_{3(23)}$	& $1.77\times10^{1}$ & ($1.83\times10^{-8}$) & $\checkmark$ &  $\checkmark$ &  $\checkmark$ &  $\times (\checkmark)$ &  $\checkmark$ &  $\checkmark$  \\ 
		$\widehat{\mathbf{B}}_{3(33)}$	& $7.61\times10^{-2}$ & ($6.27$) & $\checkmark$ &  $\checkmark$ &  $\checkmark$ &  $\checkmark (\times)$ &  $\checkmark$ &  $\checkmark$  \\ 
		$\widehat{\mathbf{B}}_{4(11)}$	& $2.32\times10^{1}$ & ($1.18\times10^{2}$) & $\checkmark$ &  $\checkmark$ &  $\checkmark(\times)$ &  $\times$ &  $\checkmark$ &  $\checkmark(\times)$  \\
		$\widehat{\mathbf{B}}_{4(12)}$	& $1.88\times10^{-2}$ & ($6.85$) & $\checkmark$ &  $\checkmark$ &  $\checkmark$ &  $\checkmark (\times)$ &  $\checkmark$ &  $\checkmark$  \\
		$\widehat{\mathbf{B}}_{4(22)}$	& $9.72\times10^{-1}$ & ($8.36\times10^{-7}$) & $\checkmark$ &  $\checkmark$ &  $\checkmark$ &  $\checkmark$ &  $\checkmark$ &  $\checkmark$  \\ 
		$\widehat{\mathbf{B}}_{4(23)}$	& $1.44\times10^{2}$ & ($1.25\times10^{-8}$) & $\checkmark$ &  $\checkmark$ &  $\checkmark$ &  $\times (\checkmark)$ &  $\checkmark$ &  $\checkmark$  \\ 
		\hline
	\end{tabular} 
	\caption{\label{tab:hybridB}The minimum of $\chi^2$ for the $\widehat{\mathbf{B}}$-type hybrid textures.}
\end{table*}

\begin{table*}[ht]
	\begin{tabular}{clccccccc}
		\hline
		Majorana $\mathbf{m}_\nu$ 	&\;\;$\chi^2_{min}$ NO &\;\;\;\; (IO)\;\;\;\;  &\; $\Delta m^2_{21}$\;  &\; $\Delta m^2_{31}$\;  & \;\;\;$\theta_{12}$\;\;\; &\;\;\;$\theta_{23}$\;\;\;  &\;\;\;$\theta_{13}$\;\;\;  &\; \;\;$\delta$\;\;\; \\ \hline	
		$\widehat{\mathbf{C}}_{(11)}$	& $1.40\times10^{-1}$ & ($1.08\times10^{2}$) & $\checkmark$ &  $\checkmark$ &  $\checkmark (\times)$ &  $\checkmark$ &  $\checkmark$ &  $\checkmark$  \\ 
		$\widehat{\mathbf{C}}_{(12)}$	& $3.19\times10^{-1}$ & ($3.68\times10^{-6}$) & $\checkmark$ &  $\checkmark$ &  $\checkmark$ &  $\checkmark$ &  $\checkmark$ &  $\checkmark$  \\ 
		$\widehat{\mathbf{C}}_{(13)}$	& $1.88\times10^{-6}$ & ($3.72$) & $\checkmark$ &  $\checkmark$ &  $\checkmark $ &  $\checkmark(\times)$ &  $\checkmark$ &  $\checkmark$  \\ 
		$\widehat{\mathbf{C}}_{(23)}$	& $6.20\times10^{2}$ & ($3.52\times10^{-11}$) & $\checkmark$ &  $\times(\checkmark)$ &  $\checkmark$ &  $\checkmark$ &  $\times(\checkmark)$ &  $\checkmark$  \\ 
		\hline
	\end{tabular} 
	\caption{\label{tab:hybridC}The minimum of $\chi^2$ for the $\widehat{\mathbf{C}}$-type hybrid textures.}
\end{table*}

\begin{table*}[ht]
	\begin{tabular}{clccccccc}
		\hline
		Majorana $\mathbf{m}_\nu$ 	&\;\;$\chi^2_{min}$ NO &\;\;\;\; (IO)\;\;\;\;  &\; $\Delta m^2_{21}$\;  &\; $\Delta m^2_{31}$\;  & \;\;\;$\theta_{12}$\;\;\; &\;\;\;$\theta_{23}$\;\;\;  &\;\;\;$\theta_{13}$\;\;\;  &\; \;\;$\delta$\;\;\; \\ \hline	
		$\widehat{\mathbf{D}}_{1(11)}$	&  $1.07\times10^{-8}$ & ($1.08\times10^{2}$) & $\checkmark$ &$\checkmark$  & $\checkmark (\times)$ & $\checkmark$ & $\checkmark$ & $\checkmark$ \\ 
		$\widehat{\mathbf{D}}_{1(12)}$	& $4.77\times10^{-8}$ & ($2.11\times10^{-6}$) & $\checkmark$ &$\checkmark$  & $\checkmark$ & $\checkmark$ & $\checkmark$ & $\checkmark$\\ 
		$\widehat{\mathbf{D}}_{1(13)}$	&  $1.53\times10^{-7}$ & ($7.88\times10^{-6}$) & $\checkmark$ &  $\checkmark$ &  $\checkmark$ &  $\checkmark$ &  $\checkmark$ &  $\checkmark$ \\ 
		$\widehat{\mathbf{D}}_{1(33)}$	& $6.88$ &($1.87$) & $\checkmark$ &  $\checkmark$ &  $\checkmark$ &  $\times ($\checkmark$)$ &  $\checkmark$ &  $\checkmark$ \\ 
		$\widehat{\mathbf{D}}_{2(11)}$	& $1.14\times10^{-8}$ & ($1.08\times10^{2}$)  & $\checkmark$ &  $\checkmark$ &  $\checkmark (\times)$ &  $\checkmark$ &  $\checkmark$ &  $\checkmark$ \\ 
		$\widehat{\mathbf{D}}_{2(12)}$	& $6.60\times10^{-10}$ & ($1.62\times10^{-6}$) & $\checkmark$ &  $\checkmark$ &  $\checkmark$ &  $\checkmark$ &  $\checkmark$ &  $\checkmark$ \\ 
		$\widehat{\mathbf{D}}_{2(13)}$	& $1.63\times10^{-7}$ & ($2.59\times10^{-7}$) & $\checkmark$ &  $\checkmark$ &  $\checkmark$ &  $\checkmark$ &  $\checkmark$ &  $\checkmark$  \\ 
		$\widehat{\mathbf{D}}_{2(22)}$	& $3.12$ & ($3.38\times10^{-3}$) & $\checkmark$ &  $\checkmark$ &  $\checkmark$ &  $\times (\checkmark)$ &  $\checkmark$ &  $\checkmark$ \\ 
		\hline
	\end{tabular} 
	\caption{\label{tab:hybridD}The minimum of $\chi^2$ for the $\widehat{\mathbf{D}}$-type hybrid textures.}
\end{table*}

\begin{table*}[ht]
	\begin{tabular}{clccccccc}
		\hline
		Majorana $\mathbf{m}_\nu$ 	&\;\;$\chi^2_{min}$ NO &\;\;\;\; (IO)\;\;\;\;  &\; $\Delta m^2_{21}$\;  &\; $\Delta m^2_{31}$\;  & \;\;\;$\theta_{12}$\;\;\; &\;\;\;$\theta_{23}$\;\;\;  &\;\;\;$\theta_{13}$\;\;\;  &\; \;\;$\delta$\;\;\; \\ \hline
		$\widehat{\mathbf{E}}_{1(12)}$	& $1.65\times10^{-6}$ & ($1.51\times10^{1}$) & $\checkmark$ &  $\checkmark (\times)$ &  $\checkmark$ &  $ \checkmark(\times)$ &  $\checkmark(\times)$ &  $\checkmark$\\
		$\widehat{\mathbf{E}}_{1(13)}$	& $7.10\times10^{-7}$ & ($1.97\times10^{2}$) & $\checkmark$ &  $\checkmark$ &  $\checkmark$ &  $\checkmark(\times)$ &  $\checkmark$ &  $\checkmark$\\	
		$\widehat{\mathbf{E}}_{1(23)}$	& $2.03\times10^{1}$ & ($1.33\times10^{-9}$) & $\checkmark$ &  $\checkmark$ &  $\checkmark$ &  $\times (\checkmark)$ &  $\checkmark$ &  $\checkmark$\\
		$\widehat{\mathbf{E}}_{1(33)}$	& $1.40$ & ($1.39\times10^{-7}$) & $\checkmark$ &  $\checkmark$ &  $\checkmark$ &  $\times (\checkmark)$ &  $\checkmark$ &  $\checkmark$\\
		$\widehat{\mathbf{E}}_{2(12)}$	& $2.55\times10^{-6}$ & ($2.58\times10^{2}$) & $\checkmark$ &  $\checkmark$ &  $\checkmark$ &  $\checkmark(\times)$ &  $\checkmark$ &  $\checkmark$\\
		$\widehat{\mathbf{E}}_{2(13)}$	& $7.41\times10^{-9}$ & ($9.09$) & $\checkmark$ &  $\checkmark(\times)$ &  $\checkmark$ &  $\checkmark(\times)$ &  $\checkmark(\times)$ &  $\checkmark$\\
		$\widehat{\mathbf{E}}_{2(22)}$	& $1.88$ & ($9.94\times10^{-1}$) & $\checkmark$ &  $\checkmark$ &  $\checkmark$ &  $\times (\checkmark)$ &  $\checkmark$ &  $\checkmark$\\
		$\widehat{\mathbf{E}}_{2(23)}$	& $1.77\times10^{2}$ & ($1.33\times10^{-9}$) & $\checkmark$ &  $\checkmark$ &  $\checkmark$ &  $\times (\checkmark)$ &  $\checkmark$ &  $\checkmark$\\	
		$\widehat{\mathbf{E}}_{3(12)}$	& $2.90\times10^{-8}$ & ($3.24\times10^{2}$) & $\checkmark$ &  $\checkmark$ &  $\checkmark(\times)$ &  $\checkmark(\times)$ &  $\checkmark$ &  $\checkmark(\times)$\\	
		$\widehat{\mathbf{E}}_{3(13)}$	& $1.52\times10^{-6}$ & ($2.51\times10^{2}$) & $\checkmark$ &  $\checkmark$ &  $\checkmark(\times)$ &  $\checkmark(\times)$ &  $\checkmark$ &  $\checkmark(\times)$\\	
		$\widehat{\mathbf{E}}_{3(22)}$	& $1.32\times10^{2}$ & ($3.17\times10^{-9}$) & $\checkmark$ &  $\checkmark$ &  $\times (\checkmark)$ &  $\times (\checkmark)$ &  $\checkmark$ &  $\checkmark$\\	
		$\widehat{\mathbf{E}}_{3(33)}$	& $2.80\times10^{2}$ & ($5.74\times10^{-4}$) & $\checkmark$ &  $\checkmark$ &  $\times (\checkmark)$ &  $\times (\checkmark)$ &  $\checkmark$ &  $\checkmark$\\					
		\hline
	\end{tabular} 
	\caption{\label{tab:hybridE}The minimum of $\chi^2$ for the $\widehat{\mathbf{E}}$-type hybrid textures.}
\end{table*}

\begin{table*}[ht]
	\begin{tabular}{clccccccc}
		\hline
		Majorana $\mathbf{m}_\nu$ &\;\;$\chi^2_{min}$ NO &\;\;\;\; (IO)\;\;\;\;  &\; $\Delta m^2_{21}$\;  &\; $\Delta m^2_{31}$\;  & \;\;\;$\theta_{12}$\;\;\; &\;\;\;$\theta_{23}$\;\;\;  &\;\;\;$\theta_{13}$\;\;\;  &\; \;\;$\delta$\;\;\; \\ \hline
		$\widehat{\mathbf{F}}_{1(11)}$	& $2.15\times10^{-12}$ & ($1.08\times10^{2}$) & $\checkmark$ &  $\checkmark$ &  $\checkmark(\times)$ &  $\checkmark$ &  $\checkmark$ &  $\checkmark$\\	
		$\widehat{\mathbf{F}}_{1(22)}$	& $8.26\times10^{-1}$ & ($3.25\times10^{-6}$) & $\checkmark$ &  $\checkmark$ &  $\checkmark$ &  $\checkmark$ &  $\checkmark$ &  $\checkmark$\\		
		$\widehat{\mathbf{F}}_{1(23)}$	& $1.88\times10^{1}$ & ($2.15\times10^{-8}$) & $\checkmark$ &  $\checkmark$ &  $\checkmark$ &  $\times(\checkmark)$ &  $\checkmark$ &  $\checkmark$\\
		$\widehat{\mathbf{F}}_{1(33)}$	& $2.87\times10^{-6}$ & ($6.97$) & $\checkmark$ &  $\checkmark$ &  $\checkmark$ &  $\checkmark (\times)$ &  $\checkmark$ &  $\checkmark$\\		
		$\widehat{\mathbf{F}}_{2(11)}$	& $1.68\times10^{1}$ & ($3.68\times10^{2}$) & $\checkmark$ &  $\checkmark$ &  $\checkmark(\times)$ &  $\times$ &  $\checkmark(\times)$ &  $\checkmark$\\	
		$\widehat{\mathbf{F}}_{2(13)}$	& $7.95\times10^{-7}$ & ($1.43\times10^{-6}$) & $\checkmark$ &  $\checkmark$ &  $\checkmark$ &  $\checkmark$ &  $\checkmark$ &  $\checkmark$\\
		$\widehat{\mathbf{F}}_{2(22)}$	& $2.87$ & ($3.62\times10^{-7}$) & $\checkmark$ &  $\checkmark$ &  $\checkmark$ &  $\times (\checkmark)$ &  $\checkmark$ &  $\checkmark$\\	
		$\widehat{\mathbf{F}}_{2(33)}$	& $2.50\times10^{2}$ & ($1.42\times10^{-6}$) & $\checkmark$ &  $\checkmark$ &  $\times (\checkmark)$ &  $\times (\checkmark)$ &  $\times (\checkmark)$ &  $\checkmark$\\
		$\widehat{\mathbf{F}}_{3(11)}$	& $1.86\times10^{1}$ & ($3.07\times10^{2}$) & $\checkmark$ &  $\checkmark$ &  $\checkmark(\times)$ &  $\times$ &  $\checkmark$ &  $\checkmark$\\	
		$\widehat{\mathbf{F}}_{3(12)}$	& $6.88\times10^{-8}$ & ($9.89\times10^{-7}$) & $\checkmark$ &  $\checkmark$ &  $\checkmark$ &  $\checkmark$ &  $\checkmark$ &  $\checkmark$\\
		$\widehat{\mathbf{F}}_{3(22)}$	& $1.04\times10^{2}$ & ($2.94\times10^{-6}$) & $\checkmark$ &  $\checkmark$ &  $\times (\checkmark)$ &  $\times (\checkmark)$ &  $\times (\checkmark)$ &  $\checkmark$\\	
		$\widehat{\mathbf{F}}_{3(33)}$	& $5.61$ & ($4.32\times10^{-6}$) & $\checkmark$ &  $\checkmark$ &  $\checkmark$ &  $\times (\checkmark)$ &  $\checkmark$ &  $\checkmark$\\							
		\hline
	\end{tabular} 
	\caption{\label{tab:hybridF}The minimum of $\chi^2$ for the $\widehat{\mathbf{F}}$-type hybrid textures.}
\end{table*}

\begin{table*}[ht]
	\begin{tabular}{clccccccc}
		\hline
		Dirac $\mathbf{m}_\nu$	&\;\;$\chi^2_{min}$ NO &\;\;\;\; (IO)\;\;\;\;  &\; $\Delta m^2_{21}$\;  &\; $\Delta m^2_{31}$\;  & \;\;\;$\theta_{12}$\;\;\; &\;\;\;$\theta_{23}$\;\;\;  &\;\;\;$\theta_{13}$\;\;\;  &\; \;\;$\delta$\;\;\; \\ \hline	
		$\widehat{\mathbf{A}}_{1(22)}$	& $1.16\times10^{1}$ & ($3.15\times10^{1}$) & $\checkmark$ &$\checkmark$  & $\checkmark(\times)$ & $\times$ & $\checkmark$ & $\checkmark$\\ 
		$\widehat{\mathbf{A}}_{1(33)}$	& $1.31\times10^{1}$ &($3.09\times10^{1}$) & $\checkmark$ &  $\checkmark$ &  $\checkmark(\times)$ &  $\times$ &  $\checkmark$ &  $\checkmark$  \\ 
		$\widehat{\mathbf{A}}_{2(22)}$	& $5.65$ & ($4.35\times10^{1}$) & $\checkmark$ &  $\checkmark$ &  $\checkmark(\times)$ &  $\times$ &  $\checkmark$ &  $\checkmark$ \\ 
		$\widehat{\mathbf{A}}_{2(33)}$	& $7.65\times10^{1}$ & ($5.01\times10^{1}$) & $\checkmark$ &  $\checkmark$ &  $\times$ &  $\times$ &  $\checkmark$ &  $\checkmark$  \\ 
		$\widehat{\mathbf{B}}_{1(11)}$	& $6.91$  &  ($3.15\times10^{2}$) & $\checkmark$ & $\checkmark$  & $\checkmark(\times)$  & $\times$  & $\checkmark$  &  $\checkmark(\times)$ \\
		$\widehat{\mathbf{B}}_{1(33)}$	&  $1.72\times10^{-2}$ & ($5.40\times10^{1}$) & $\checkmark$ &  $\checkmark(\times)$ &  $\checkmark$ &  $\checkmark(\times)$ &  $\checkmark(\times)$ &  $\checkmark(\times)$ \\ 
		$\widehat{\mathbf{B}}_{2(11)}$	& $3.07\times10^{1}$  & ($3.80\times10^{2}$) & \checkmark & \checkmark  & $\checkmark(\times)$  & $\times$  & $\checkmark$  & $\checkmark(\times)$  \\
		$\widehat{\mathbf{B}}_{2(22)}$	& $9.05\times10^{-1}$ & ($4.16\times10^{1}$)  & $\checkmark$ &  $\checkmark(\times)$ &  $\checkmark$ &  $\checkmark$ &  $\checkmark(\times)$ &  $\checkmark$ \\ 
		$\widehat{\mathbf{B}}_{3(11)}$	& $6.37$  & ($1.08\times10^{2}$) & $\checkmark$ & $\checkmark$  & $\checkmark(\times)$  & $\times(\checkmark)$  & $\checkmark$  & $\checkmark$  \\
		$\widehat{\mathbf{B}}_{3(33)}$	& $6.56\times10^{-3}$ & ($1.41\times10^{1}$) & $\checkmark$ &  $\checkmark$ &  $\checkmark(\times)$ &  $\checkmark(\times)$ &  $\checkmark$ &  $\checkmark(\times)$  \\ 
		$\widehat{\mathbf{B}}_{4(11)}$	&  $2.62\times10^{1}$  & ( $1.18\times10^{2}$) &  $\checkmark$ &  $\checkmark$  &  $\checkmark(\times)$  & $\times$  &  $\checkmark$  &  $\checkmark(\times)$  \\
		$\widehat{\mathbf{B}}_{4(22)}$	& $9.89\times10^{-1}$ & ($3.43\times10^{1}$) & $\checkmark$ &  $\checkmark$ &  $\checkmark(\times)$ &  $\checkmark(\times)$ &  $\checkmark$ &  $\checkmark(\times)$  \\ 
		$\widehat{\mathbf{C}}_{(11)}$	& $2.51\times10^{-1}$ & ($1.08\times10^{2}$) & $\checkmark$ &  $\checkmark$ &  $\checkmark (\times)$ &  $\checkmark$ &  $\checkmark$ &  $\checkmark$  \\ 
		$\widehat{\mathbf{D}}_{1(11)}$	&  $4.51\times10^{-1}$ & ($1.08\times10^{2}$) & $\checkmark$ &$\checkmark$  & $\checkmark (\times)$ & $\checkmark$ & $\checkmark$ & $\checkmark$ \\ 
		$\widehat{\mathbf{D}}_{1(33)}$	& $1.60\times10^{1}$ &($2.62$) & $\checkmark$ &  $\checkmark$ &  $\times(\checkmark)$ &  $\times ($\checkmark$)$ &  $\checkmark$ &  $\times$ \\ 
		$\widehat{\mathbf{D}}_{2(11)}$	& $1.49$ & ($1.08\times10^{2}$)  & $\checkmark$ &  $\checkmark$ &  $\checkmark (\times)$ &  $\times(\checkmark)$ &  $\checkmark$ &  $\checkmark$ \\ 
		$\widehat{\mathbf{D}}_{2(22)}$	& $5.47$ & ($5.88\times10^{-1}$) & $\checkmark$ &  $\checkmark$ &  $\checkmark$ &  $\times (\checkmark)$ &  $\checkmark$ &  $\checkmark$ \\ 
		$\widehat{\mathbf{E}}_{1(33)}$	& $5.76\times10^{1}$ & ($7.34\times10^{-1}$) & $\checkmark$ &  $\checkmark$ &  $\times(\checkmark)$ &  $\times(\checkmark)$ &  $\checkmark$ &  $\checkmark$\\
		$\widehat{\mathbf{E}}_{2(22)}$	& $5.79\times10^{1}$ & ($8.78$) & $\checkmark$ &  $\checkmark$ &  $\times$ &  $\times $ &  $\checkmark$ &  $\checkmark(\times)$\\		
		$\widehat{\mathbf{E}}_{3(22)}$	& $1.33\times10^{2}$ & ($1.92\times10^{-3}$) & $\checkmark$ &  $\checkmark$ &  $\times (\checkmark)$ &  $\times (\checkmark)$ &  $\checkmark$ &  $\times(\checkmark)$\\	
		$\widehat{\mathbf{E}}_{3(33)}$	& $2.80\times10^{2}$ & ($4.02\times10^{-1}$) & $\checkmark$ &  $\checkmark$ &  $\times (\checkmark)$ &  $\times (\checkmark)$ &  $\checkmark$ &  $\times(\checkmark)$\\
		$\widehat{\mathbf{F}}_{1(11)}$	& $2.94\times10^{-2}$  & ($1.08\times10^{2}$) & $\checkmark$ & $\checkmark$  & $\checkmark(\times)$  & $\checkmark$  & $\checkmark$  & $\checkmark$  \\	
		$\widehat{\mathbf{F}}_{1(22)}$	& $1.37\times10^{1}$ & ($2.12$) & $\checkmark$ &  $\checkmark$ &  $\checkmark$ &  $\times(\checkmark)$ &  $\checkmark$ &  $\checkmark (\times)$\\
		$\widehat{\mathbf{F}}_{1(33)}$	& $9.53\times10^{1}$ & ($3.50$) & $\checkmark$ &  $\checkmark$ &  $\checkmark$ &  $\times$ &  $\checkmark$ &  $\times(\checkmark)$\\	
		$\widehat{\mathbf{F}}_{2(11)}$	& $2.00\times10^{1}$  & ($3.68\times10^{2}$) & $\checkmark$ & $\checkmark$  & $\checkmark(\times)$  & $\times$  & $\checkmark(\times)$  &  $\checkmark$ \\	
		$\widehat{\mathbf{F}}_{2(22)}$	& $9.31$ & ($2.40\times10^{1}$) & $\checkmark$ &  $\checkmark$ &  $\checkmark(\times)$ &  $\times (\checkmark)$ &  $\checkmark$ &  $\checkmark(\times)$\\	
		$\widehat{\mathbf{F}}_{2(33)}$	& $2.50\times10^{2}$ & ($1.34$) & $\checkmark$ &  $\checkmark$ &  $\times (\checkmark)$ &  $\times (\checkmark)$ &  $\times (\checkmark)$ &  $\checkmark$\\
		$\widehat{\mathbf{F}}_{3(11)}$	& $1.86\times10^{1}$  & ($3.07\times10^{2}$) & $\checkmark$ & $\checkmark$  & $\checkmark(\times)$  & $\times$  & $\checkmark$  & $\checkmark$  \\
		$\widehat{\mathbf{F}}_{3(22)}$	& $1.04\times10^{2}$ & ($1.22\times10^{1}$) & $\checkmark$ &  $\checkmark$ &  $\times$ &  $\times$ &  $\times (\checkmark)$ &  $\times$\\	
		$\widehat{\mathbf{F}}_{3(33)}$	& $1.95\times10^{1}$ & ($2.37\times10^{1}$) & $\checkmark$ &  $\checkmark$ &  $\times$ &  $\times (\checkmark)$ &  $\checkmark$ &  $\checkmark$\\				
		\hline
	\end{tabular} 
	\caption{\label{tab:hybrid-Dirac}The minimum of $\chi^2$ for Dirac-type hybrid textures. We present only the patterns for which the Dirac phase $\delta$ is different from 0 or $\pi$ and CP is violated.}
\end{table*}

Hybrid textures~\cite{Kaneko:2005yz} are particular cases of one-zero textures of the Majorana neutrino mass matrix, which additionally have two equal nonzero elements, and are defined in the flavor basis. There are $(6!/5!) \times 5!/(2!\, 3!)=60$ possible hybrid textures. Among them, it has been shown that only 39 textures are compatible with current neutrino oscillation data at the $3\sigma$ level~\cite{Liu:2013oxa}. To keep a coherent notation, without the need of introducing any new classification scheme, we shall label these matrices as follows. We associate to each FGM matrix $\mathbf{M}$ given in Eq.~\eqref{FGMtextures} a hybrid-type matrix $\widehat{\mathbf{M}}$, in which the two zeros in $\mathbf{M}$ are replaced by equal nonvanishing elements in $\widehat{\mathbf{M}}$. Then, the position of the zero element in the hybrid matrix $\widehat{\mathbf{M}}$ is indicated with a subscript in parenthesis. Consider, for instance, the hybrid texture
\begin{align}
\label{hybridexample}
\begin{pmatrix}
\text{X} & \text{X} & \ast \\ \text{X} &  \ast &  \ast \\ \ast & \ast & 0
\end{pmatrix}\,,
\end{align}
where ``X" stands for equal nonzero elements. Following the definition of the matrix $\mathbf{A}_1$ given in Eq.~\eqref{FGMtextures}, the hybrid matrix~\eqref{hybridexample} would be represented as $\widehat{\mathbf{A}}_{1(33)}$ in our notation. Obviously, for each FGM texture in Eq.~\eqref{FGMtextures}, one can construct four different hybrid textures, depending on the position of the zero matrix element.

For comparison, below we list the complete set of 39 hybrid textures studied in Ref.~\cite{Liu:2013oxa}:
\begin{align}
	\label{hybridtextures}
	&\widehat{\mathbf{A}}_{1\,\{(13),(22),(23),(33)\}}\,, 
	~\widehat{\mathbf{A}}_{2\,\{(12),(22),(23),(33)\}}\,, \nonumber\\
    &\widehat{\mathbf{B}}_{1\,\{(12),(23),(33)\}}\,,
    \widehat{\mathbf{B}}_{2\,\{(13),(22),(23)\}}\,,\nonumber\\
   &\widehat{\mathbf{B}}_{3\,\{(13),(23),(33)\}}\,,
    ~\widehat{\mathbf{B}}_{4\,\{(12),(22),(23)\}}\,,
    ~\widehat{\mathbf{C}}_{(11)}\,,\nonumber\\
    &\widehat{\mathbf{D}}_{1\,\{(11),(12),(13),(33)\}}\,,
    ~\widehat{\mathbf{D}}_{2\,\{(11),(12),(13),(22)\}}\,,\\ &\widehat{\mathbf{E}}_{1(33)}\,,~\widehat{\mathbf{E}}_{2(22)}\,,
    ~\widehat{\mathbf{E}}_{3\,\{(22),(33)\}}\,,\nonumber\\ &\widehat{\mathbf{F}}_{1\,\{(22),(33)\}}\,,
    ~\widehat{\mathbf{F}}_{2\,\{(22),(33)\}}\,,
    ~\widehat{\mathbf{F}}_{3\,\{(22),(33)\}}\,,\nonumber
\end{align}
where we have indicated, inside curly brackets, the possible choices for the texture-zero position.

The results of the $\chi^2$-minimization are summarized in Tables~\ref{tab:hybridA}-\ref{tab:hybridF}. First we note that all textures given in Eq.~\eqref{hybridtextures} are compatible with data at the $1\sigma$ level either for NO, IO or both types of neutrino mass spectrum. In particular, the patterns $\widehat{\mathbf{A}}_{2(33)}$, $\widehat{\mathbf{B}}_{1(12)}$, $\widehat{\mathbf{B}}_{2(22)}$, $\widehat{\mathbf{B}}_{3(13)}$, $\widehat{\mathbf{B}}_{4(22)}$, $\widehat{\mathbf{C}}_{(12)}$, $\widehat{\mathbf{D}}_{1(12)}$, $\widehat{\mathbf{D}}_{1(13)}$, $\widehat{\mathbf{D}}_{2(12)}$, $\widehat{\mathbf{D}}_{2(13)}$, $\widehat{\mathbf{F}}_{1(22)}$, $\widehat{\mathbf{F}}_{2(13)}$, and $\widehat{\mathbf{F}}_{3(12)}$ turn out to be compatible with experimental data for NO and IO mass spectra.

Despite the fact that our analysis is performed at the stringent $1\sigma$ C.L., constraining also the Dirac phase $\delta$, among the sixty possible hybrid patterns for Majorana neutrinos, only six fail in reproducing the data for any hierarchy and can be completely excluded. These are the matrices $\widehat{\mathbf{B}}_{1(11)}$, $\widehat{\mathbf{B}}_{2(11)}$, $\widehat{\mathbf{B}}_{3(11)}$, $\widehat{\mathbf{B}}_{4(11)}$, $\widehat{\mathbf{F}}_{2(11)}$, and $\widehat{\mathbf{F}}_{3(11)}$, all having a zero element in the (1,1) position. 

We remark that in Ref.~\cite{Liu:2013oxa}, only 13 patterns were found compatible with data at the $1\sigma$ level: $\widehat{\mathbf{A}}_{1(22)}$, $\widehat{\mathbf{A}}_{1(23)}$, $\widehat{\mathbf{A}}_{1(33)}$, $\widehat{\mathbf{B}}_{1(23)}$, $\widehat{\mathbf{B}}_{1(33)}$, $\widehat{\mathbf{B}}_{2(22)}$, $\widehat{\mathbf{B}}_{3(13)}$, $\widehat{\mathbf{B}}_{4(23)}$, $\widehat{\mathbf{D}}_{2(11)}$, $\widehat{\mathbf{D}}_{2(13)}$, $\widehat{\mathbf{E}}_{2(22)}$, $\widehat{\mathbf{F}}_{2(22)}$, and $\widehat{\mathbf{F}}_{2(33)}$. The fact that several viable hybrid textures were missed in Ref.~\cite{Liu:2013oxa} could be attributed to the numerical procedure followed by the authors, who performed a simple random scanning of the parameter space instead of the more reliable $\chi^2$-analysis.

In the case of Dirac neutrinos, thirty patterns were considered, which are listed in Table~\ref{tab:hybrid-Dirac}. We include all the Hermitian patterns that do not have any off-diagonal zero element, and thus may lead to Dirac-type CP violation. For the remaining patterns, the Dirac phase $\delta$ is always 0 or $\pi$ and CP is conserved in the lepton sector. Looking at the table we note that only twelve textures are consistent with data either for NO or IO neutrino mass spectrum. These are the matrices $\widehat{\mathbf{B}}_{1(33)}$, $\widehat{\mathbf{B}}_{2(22)}$, $\widehat{\mathbf{B}}_{3(33)}$, $\widehat{\mathbf{B}}_{4(22)}$, $\widehat{\mathbf{C}}_{(11)}$, $\widehat{\mathbf{D}}_{1(11)}$, $\widehat{\mathbf{D}}_{2(22)}$, $\widehat{\mathbf{E}}_{1(33)}$, $\widehat{\mathbf{E}}_{3(22)}$, $\widehat{\mathbf{E}}_{3(33)}$, $\widehat{\mathbf{F}}_{1(11)}$, and $\widehat{\mathbf{F}}_{2(33)}$. None of these matrices is simultaneously allowed for both mass spectra.

\section{Parallel textures}
\label{sec:partextures}

In this section, we perform a systematic $\chi^2$-analysis of lepton mass matrices that exhibit the same texture, i.e. with $\mathbf{m}_\ell$ and $\mathbf{m}_\nu$ having their zeros located at the same positions. Besides the possibility of implementing a universal flavor structure in the context of grand unified models, there is an additional theoretical motivation for considering parallel structures. It is well known that an attractive and economical framework to generate small neutrino masses is the seesaw mechanism. In its simplest type-I realization, three right-handed neutrinos are added to the standard model particle content. It is then conceivable that the presence of family symmetries enforces texture-zero structures in the Dirac neutrino mass matrix $\mathbf{m}_D$ and the heavy Majorana mass matrix $\mathbf{M}_R$, which, in some cases, could be preserved by the effective neutrino mass matrix $\mathbf{m}_\nu=-\mathbf{m}_D \mathbf{M}_R^{-1} \mathbf{m}_D^\mathsf{T}$.\footnote{The patterns belonging to the classes I and IV in Eq.~\eqref{psets} have this property~\cite{Branco:2007nn}.} 

It is worth noticing that any permutation transformation acting on parallel patterns is allowed, since it leads to textures with the same physical content. Indeed, they can be related by a weak basis transformation, performed by a permutation matrix $\mathbf{P}$,
\begin{equation}
\label{wbtransf}
\mathbf{m}^{\prime}_{\ell}=\mathbf{P}^{\mathsf{T}}\,\mathbf{m}_{\ell}\,\mathbf{P}\,,\quad
\mathbf{m}^{\prime}_{\nu}=\mathbf{P}^{\mathsf{T}}\,\mathbf{m}_{\nu}\,\mathbf{P}\,,
\end{equation}
which automatically preserves the parallel structure, but changes the position of the zeros. The matrix $\mathbf{P}$ belongs to the group of six permutations matrices, which are isomorphic to the symmetry group $S_3\,$.

\subsection{Two-zero textures}

The FGM-type Ans\"atze can be classified into four weak basis equivalent classes (or permutation sets)~\cite{Branco:2007nn}:
\begin{align}\label{psets}
\begin{split}
\text{Class I:} &\quad \mathbf{A}_1, \mathbf{A}_2, \mathbf{B}_3,\mathbf{B}_4,
\mathbf{D}_1, \mathbf{D}_2;\\
\text{Class II:} &\quad \mathbf{B}_1, \mathbf{B}_2, \mathbf{E}_3; \\
\text{Class III:} &\quad \mathbf{C}, \mathbf{E}_1, \mathbf{E}_2;\\
\text{Class IV:} &\quad \mathbf{F}_1, \mathbf{F}_2, \mathbf{F}_3.
\end{split}
\end{align}
It is clear that class IV is not experimentally viable, since it always leads to the decoupling of one generation. Note also that the weak basis transformations given in Eq.~\eqref{wbtransf} are not allowed in a scheme with a diagonal and ordered charged lepton mass matrix, as in the texture schemes discussed in previous section. 

In our $\chi^2$-analysis, all parallel FGM textures with arbitrary complex Hermitian (or real symmetric) $\mathbf{m}_{\ell}$ and complex symmetric $\mathbf{m}_{\nu}$ were found to be viable for both normal and inverted neutrino mass ordering. Similar results were obtained for Dirac neutrinos with Hermitian neutrino mass matrices.  

We have also considered the feasibility of arbitrary complex Hermitian $\mathbf{m}_{\ell}$ and real symmetric $\mathbf{m}_{\nu}$. In this case, the number of physical parameters is equal to 10 for classes~I and~II, while for class~III there are 11 parameters since, in general, the invariant quantity $\arg\bigl[(\mathbf{m}_{\ell})_{12}(\mathbf{m}_{\ell}^{\ast})_{13}(\mathbf{m}_{\ell})_{23}\bigr]$ does not vanish. As far as the analysis of the neutrino oscillation data is concerned, there is no distinction between Majorana or Dirac neutrinos. The minimum of $\chi^2$ was always found to be much smaller than one, so that all patterns in classes I, II, and III are consistent with neutrino data for any mass hierarchy.

\begin{table*}[ht]
	\begin{tabular}{clccccccc}
		\hline
		Majorana $\mathbf{m}_\nu$	&\;\;$\chi^2_{min}$ NO &\;\;\;\; (IO)\;\;\;\;  &\; $\Delta m^2_{21}$\;  &\; $\Delta m^2_{31}$\;  & \;\;\;$\theta_{12}$\;\;\; &\;\;\;$\theta_{23}$\;\;\;  &\;\;\;$\theta_{13}$\;\;\;  &\; \;\;$\delta$\;\;\; \\ \hline	
		$\mathbf{A}_{1(13)}$	& $1.25\times10^{-7}$ & ($1.71\times10^{-8}$) & $\checkmark$ & $\checkmark$  & $\checkmark$ & $\checkmark$ & $\checkmark$ & $\checkmark$\\
		$\mathbf{A}_{1(22)}$	& $9.66\times10^{-8}$ &($4.95\times10^{-9}$) & $\checkmark$ &  $\checkmark$ &  $\checkmark$ &  $\checkmark$ &  $\checkmark$ &  $\checkmark$\\ 
		$\mathbf{A}_{1(23)}$	& $5.53\times10^{-8}$ & ($3.06$) & $\checkmark$ &  $\checkmark$ &  $\checkmark$ &  $\checkmark(\times)$ &  $\checkmark$ &  $\checkmark(\times)$ \\ 
		$\mathbf{A}_{1(33)}$	& $1.34\times10^{-7}$ & ($5.13\times10^{-8}$) & $\checkmark$ &  $\checkmark$ &  $\checkmark$ &  $\checkmark$ &  $\checkmark$ &  $\checkmark$\\ 
		$\mathbf{A}_{2(22)}$	& $7.02\times10^{-8}$  &  ($9.65\times10^{-8}$) & $\checkmark$ & $\checkmark$  & $\checkmark$  & $\checkmark$  & $\checkmark$  &  $\checkmark$\\
		$\mathbf{A}_{2(23)}$	&  $1.31\times10^{-7}$ & ($4.51\times10^{-6}$) & $\checkmark$ &  $\checkmark$ &  $\checkmark$ &  $\checkmark$ &  $\checkmark$ &  $\checkmark$ \\ 				
		$\mathbf{A}_{2(33)}$	& $7.88\times10^{-8}$  & ($3.38\times10^{-8}$) & $\checkmark$ & $\checkmark$  & $\checkmark$  & $\checkmark$  & $\checkmark$  & $\checkmark$\\		
		$\mathbf{B}_{1(23)}$	& $5.35\times10^{-7}$ & ($3.78\times10^{-6}$)  & $\checkmark$ &  $\checkmark$ &  $\checkmark$ &  $\checkmark$ &  $\checkmark$ &  $\checkmark$ \\ 
		$\mathbf{B}_{1(33)}$	& $1.38\times10^{-6}$  & ($4.57\times10^{-8}$) & $\checkmark$ & $\checkmark$  & $\checkmark$  & $\checkmark$  & $\checkmark$  & $\checkmark$  \\
		$\mathbf{B}_{2(13)}$	& $1.52$ & ($2.88\times10^{-7}$) & $\checkmark$ &  $\checkmark$ &  $\checkmark$ &  $\checkmark$ &  $\checkmark$ &  $\checkmark$  \\
		$\mathbf{B}_{2(22)}$	&  $7.93\times10^{-7}$  & ($1.54\times10^{-8}$) &  $\checkmark$ &  $\checkmark$  &  $\checkmark$  & $\checkmark$  &  $\checkmark$  &  $\checkmark$  \\
		$\mathbf{B}_{2(23)}$	& $1.52$ & ($6.29\times10^{-5}$) & $\checkmark$ &  $\checkmark$ &  $\checkmark$ &  $\checkmark$ &  $\checkmark$ &  $\checkmark$  \\ 
		$\mathbf{B}_{3(13)}$	& $1.84\times10^{-7}$ & ($3.05$) & $\checkmark$ &  $\checkmark$ &  $\checkmark$ &  $\checkmark(\times)$ &  $\checkmark$ &  $\checkmark(\times)$  \\ 
		$\mathbf{B}_{3(23)}$	&  $1.20\times10^{-6}$ & ($4.05\times10^{-8}$) & $\checkmark$ &$\checkmark$  & $\checkmark$ & $\checkmark$ & $\checkmark$ & $\checkmark$ \\ 
		$\mathbf{B}_{4(23)}$	& $1.95\times10^{-7}$ &($6.08\times10^{-9}$) & $\checkmark$ &  $\checkmark$ &  $\checkmark$ &  $\checkmark$ &  $\checkmark$ &  $\checkmark$ \\ 
		$\mathbf{C}_{(11)}$	& $1.49\times10^{-6}$ & ($1.75\times10^{-8}$)  & $\checkmark$ &  $\checkmark$ &  $\checkmark$ &  $\checkmark$ &  $\checkmark$ &  $\checkmark$ \\ 
		$\mathbf{C}_{(23)}$	& $2.69\times10^{-8}$ & ($2.43\times10^{-8}$) & $\checkmark$ &  $\checkmark$ &  $\checkmark$ &  $\checkmark$ &  $\checkmark$ &  $\checkmark$ \\ 
		$\mathbf{D}_{1(11)}$	& $4.99\times10^{-7}$ & ($1.61\times10^{-7}$) & $\checkmark$ &  $\checkmark$ &  $\checkmark$ &  $\checkmark$ &  $\checkmark$ &  $\checkmark$\\	
		$\mathbf{D}_{2(11)}$	& $4.21\times10^{-6}$ & ($5.29\times10^{-8}$) & $\checkmark$ &  $\checkmark$ &  $\checkmark$ &  $\checkmark $ &  $\checkmark$ &  $\checkmark$\\		
		$\mathbf{F}_{1(23)}$	& $1.98\times10^{1}$ & ($1.83\times10^{2}$) & $\checkmark$ &  $\checkmark$ &  $\checkmark (\times)$ &  $\times$ &  $\checkmark$ &  $\times$\\			
		\hline
	\end{tabular} 
	\caption{\label{tab:nniMajorana} The minimum of $\chi^2$ for three-zero Majorana neutrino textures, with an NNI charged lepton mass matrix.}
\end{table*}

\subsection{Three-zero textures}

There are only 6 possible three-zero parallel textures that can be constructed for both the charged-lepton and Majorana neutrino mass matrices. Since these matrices are related by weak basis transformations (permutations), they all have the same physical content and thus lead to the same predictions. We denote them by
\begin{align}
\label{3zerotextures}
\mathbf{A}_{1(33)}\,, ~\mathbf{A}_{2(22)}\,,~\mathbf{B}_{1(33)}\,,
~\mathbf{B}_{2(22)}\,, ~\mathbf{D}_{1(11)}\,, ~\mathbf{D}_{2(11)}\,,
\end{align}
where the subscript in parenthesis refers to the position of the additional zero in the corresponding two-zero texture given in Eq.~\eqref{FGMtextures}. Note that the matrix $\mathbf{C}_{(11)}$ is not included in the above list since it is traceless and, therefore, incompatible with the lepton masses. Furthermore, textures with null determinant or those leading to the decoupling of one generation have also been excluded. The texture $\mathbf{A}_{2(22)}$ is known as the nearest-neighbor-interaction pattern~\cite{Fritzsch:1979zq,Branco:1988iq}. In the context of the standard model, it has been shown that imposing an NNI texture simultaneously in the up- and down-quark sectors simply corresponds to a weak basis choice~\cite{Branco:1988iq}.  For non-Hermitian quark mass matrices, this is an example of parallel four-zero textures without any physical content. This is not necessarily true in the lepton sector, unless neutrinos are Dirac particles. For Majorana neutrinos, the assumption of a parallel NNI structure would have physical implications.  

For an arbitrary complex Hermitian $\mathbf{m}_{\ell}$ and a complex symmetric $\mathbf{m}_{\nu}$ (Majorana neutrinos), the above parallel 3-zero textures contain 9 physical parameters. No viable solution was found either for NO ($\chi^2_{min}\simeq 74$) or IO ($\chi^2_{min}\simeq 182$) neutrino mass spectrum. For a normal ordering of neutrino masses, all the textures fail in reproducing the three mixing angles, while for an inverted spectrum the mixing angles $\theta_{12}$ and $\theta_{23}$, and the phase $\delta$ did not satisfy the required $\chi^2$ criteria.

In Fig.~\ref{fig:fig-3zeros-Majorana} of Appendix~\ref{appendix2}, we present the probability distribution of the neutrino observables, obtained for the textures given in Eq.~\eqref{3zerotextures}, for NO and IO mass spectrum, respectively. For Dirac neutrinos, with the matrix $\mathbf{m}_{\nu}$ being Hermitian, similar results were found, thus excluding these patterns for both NO and IO mass spectra.\footnote{Our conclusions do not agree with the result of Ref.~\cite{Fritzsch:2015haa}, in which the parallel NNI texture $\mathbf{A}_{2(22)}$ is found to be compatible with the experimental data.}

\section{Predictive neutrino textures with  NNI charged lepton mass matrix}
\label{sec:nnitextures}
In the previous section, we have considered parallel structures for both lepton sectors, assuming an Hermitian charged lepton mass matrix. In particular, we showed that the parallel NNI texture $\mathbf{A}_{2(22)}$ is not compatible with the current neutrino data. In this section, we shall lift the assumption of Hermiticity on the NNI pattern of the charged lepton mass matrix and look for viable predictive neutrino zero textures. Such patterns are of interest since they contain the same number of physical parameters as the FGM and hybrid textures (assuming that $\mathbf{m}_{\nu}$ has three zeros). From the theoretical viewpoint, NNI lepton structures are also well motivated. For instance, it is possible to conceive flavour symmetries in the two-Higgs doublet model~\cite{Branco:2010tx}, in supersymmetric theories~\cite{Babu:2009nn}, and in grand unified models based on $\mathsf{SU}(5)$~\cite{EmmanuelCosta:2011jq,Emmanuel-Costa:2013gia} that lead to NNI textures.  In the latter models, the charged lepton mass matrix $\mathbf{m}_{\ell}$ exhibits an NNI pattern, while the effective neutrino mass matrix $\mathbf{m}_{\nu}$ contains some vanishing elements.

We shall assume that the non-Hermitian charged lepton mass matrix $\mathbf{m}_{\ell}$ is described by the NNI form $\mathbf{A}_{2(22)}$ and search for a maximal number of zeros in $\mathbf{m}_{\nu}$ compatible with the data. As before, we take $\mathbf{m}_{\nu}$ as a general complex symmetric matrix for Majorana neutrinos and an Hermitian matrix for Dirac neutrinos. We remark that, without loss of generality, all the non-vanishing matrix elements in $\mathbf{m}_{\ell}$ can be assumed real and positive. Thus, there remain two free parameters in $\mathbf{m}_{\ell}$ after fitting the charged lepton masses. 

In our $\chi^2$-search, none of the neutrino textures with more than three zeros was found compatible with the observed neutrino data. In Table~\ref{tab:nniMajorana} and~\ref{tab:nniDirac} we present the results for three-zero $\mathbf{m}_{\nu}$ textures for Majorana and Dirac neutrinos, respectively. As  can be seen from the tables, only the pattern $\mathbf{F}_{1(23)}$ fails in reproducing the data for any neutrino spectra. Moreover, the patterns $\mathbf{A}_{1(23)}$ and $\mathbf{B}_{3(13)}$ are compatible with the data only for NO neutrino masses. The remaining 17 textures are viable at the $1\sigma$ level irrespective of the mass ordering. In particular, once the Hermiticity of $\mathbf{m}_{\ell}$ is lifted, the parallel structure $\mathbf{A}_{2(22)}$ turns out to be consistent with data. We remark that taking an NNI Hermitian $\mathbf{m}_{\ell}$ together with any non-parallel three-zero neutrino pattern does not lead to  viable pairs of textures. Therefore, the non-Hermiticity condition of the charged lepton mass matrix is a crucial ingredient in this particular case.
 
\begin{table*}[ht]
	\begin{tabular}{clccccccc}
		\hline
		Dirac $\mathbf{m}_\nu$	&\;\;$\chi^2_{min}$ NO &\;\;\;\; (IO)\;\;\;\;  &\; $\Delta m^2_{21}$\;  &\; $\Delta m^2_{31}$\;  & \;\;\;$\theta_{12}$\;\;\; &\;\;\;$\theta_{23}$\;\;\;  &\;\;\;$\theta_{13}$\;\;\;  &\; \;\;$\delta$\;\;\; \\ \hline	
		$\mathbf{A}_{1(13)}$	& $6.44\times10^{-7}$ & ($3.51\times10^{-8}$) & $\checkmark$ & $\checkmark$  & $\checkmark$ & $\checkmark$ & $\checkmark$ & $\checkmark$\\
		$\mathbf{A}_{1(22)}$	& $3.20\times10^{-8}$ &($1.93\times10^{-8}$) & $\checkmark$ &  $\checkmark$ &  $\checkmark$ &  $\checkmark$ &  $\checkmark$ &  $\checkmark$\\ 
		$\mathbf{A}_{1(23)}$	& $1.18\times10^{-8}$ & ($3.06$) & $\checkmark$ &  $\checkmark$ &  $\checkmark$ &  $\checkmark(\times)$ &  $\checkmark$ &  $\checkmark(\times)$ \\ 
		$\mathbf{A}_{1(33)}$	& $5.08\times10^{-7}$ & ($6.03\times10^{-8}$) & $\checkmark$ &  $\checkmark$ &  $\checkmark$ &  $\checkmark$ &  $\checkmark$ &  $\checkmark$\\ 
		$\mathbf{A}_{2(22)}$	& $5.44\times10^{-8}$  &  ($2.25\times10^{-7}$) & $\checkmark$ & $\checkmark$  & $\checkmark$  & $\checkmark$  & $\checkmark$  &  $\checkmark$\\
		$\mathbf{A}_{2(23)}$	&  $1.46\times10^{-7}$ & ($6.74\times10^{-7}$) & $\checkmark$ &  $\checkmark$ &  $\checkmark$ &  $\checkmark$ &  $\checkmark$ &  $\checkmark$ \\ 				
		$\mathbf{A}_{2(33)}$	& $2.36\times10^{-7}$  & ($6.21\times10^{-8}$) & $\checkmark$ & $\checkmark$  & $\checkmark$  & $\checkmark$  & $\checkmark$  & $\checkmark$\\		
		$\mathbf{B}_{1(23)}$	& $1.31\times10^{-7}$ & ($5.63\times10^{-3}$)  & $\checkmark$ &  $\checkmark$ &  $\checkmark$ &  $\checkmark$ &  $\checkmark$ &  $\checkmark$ \\ 
		$\mathbf{B}_{1(33)}$	& $2.01\times10^{-5}$  & ($1.27\times10^{-7}$) & $\checkmark$ & $\checkmark$  & $\checkmark$  & $\checkmark$  & $\checkmark$  & $\checkmark$  \\
		$\mathbf{B}_{2(13)}$	& $1.52$ & ($1.51\times10^{-1}$) & $\checkmark$ &  $\checkmark$ &  $\checkmark$ &  $\checkmark$ &  $\checkmark$ &  $\checkmark$  \\
		$\mathbf{B}_{2(22)}$	&  $9.22\times10^{-7}$  & ($1.64\times10^{-8}$) &  $\checkmark$ &  $\checkmark$  &  $\checkmark$  & $\checkmark$  &  $\checkmark$  &  $\checkmark$  \\
		$\mathbf{B}_{2(23)}$	& $1.52$ & ($1.28\times10^{-6}$) & $\checkmark$ &  $\checkmark$ &  $\checkmark$ &  $\checkmark$ &  $\checkmark$ &  $\checkmark$  \\ 
		$\mathbf{B}_{3(13)}$	& $1.43\times10^{-6}$ & ($3.05$) & $\checkmark$ &  $\checkmark$ &  $\checkmark$ &  $\checkmark(\times)$ &  $\checkmark$ &  $\checkmark(\times)$  \\ 
		$\mathbf{B}_{3(23)}$	&  $7.99\times10^{-7}$ & ($9.56\times10^{-9}$) & $\checkmark$ &$\checkmark$  & $\checkmark$ & $\checkmark$ & $\checkmark$ & $\checkmark$ \\ 
		$\mathbf{B}_{4(23)}$	& $2.83\times10^{-7}$ &($6.08\times10^{-8}$) & $\checkmark$ &  $\checkmark$ &  $\checkmark$ &  $\checkmark$ &  $\checkmark$ &  $\checkmark$ \\ 
		$\mathbf{C}_{(11)}$	& $5.62\times10^{-7}$ & ($6.20\times10^{-8}$)  & $\checkmark$ &  $\checkmark$ &  $\checkmark$ &  $\checkmark$ &  $\checkmark$ &  $\checkmark$ \\ 
		$\mathbf{C}_{(23)}$	& $9.84\times10^{-8}$ & ($1.90\times10^{-8}$) & $\checkmark$ &  $\checkmark$ &  $\checkmark$ &  $\checkmark$ &  $\checkmark$ &  $\checkmark$ \\ 
		$\mathbf{D}_{1(11)}$	& $1.23\times10^{-6}$ & ($1.28\times10^{-7}$) & $\checkmark$ &  $\checkmark$ &  $\checkmark$ &  $\checkmark$ &  $\checkmark$ &  $\checkmark$\\	
		$\mathbf{D}_{2(11)}$	& $4.23\times10^{-6}$ & ($2.14\times10^{-8}$) & $\checkmark$ &  $\checkmark$ &  $\checkmark$ &  $\checkmark $ &  $\checkmark$ &  $\checkmark$\\		
		$\mathbf{F}_{1(23)}$	& $1.98\times10^{1}$ & ($1.83\times10^{2}$) & $\checkmark$ &  $\checkmark$ &  $\checkmark (\times)$ &  $\times$ &  $\checkmark$ &  $\times$\\			
		\hline
	\end{tabular} 
	\caption{\label{tab:nniDirac} The minimum of $\chi^2$ for three-zero Dirac neutrino textures, with an NNI charged lepton mass matrix.}
\end{table*}

\section{Conclusions}
\label{sec:summary}

There has been lately a revival of the interest in texture-zero models that aim at explaining the flavor structure observed in lepton mass matrices. In this work, we have confronted various popular texture-zero \emph{Ans\"{a}tze} of lepton mass matrices with current neutrino data. We have performed a thorough $\chi^2$-analysis in a wide class of schemes, considering Hermitian charged lepton mass matrices in combination with symmetric Majorana or Hermitian Dirac neutrino mass matrices. In our study we included the well-known FGM textures, the so-called hybrid textures, as well as parallel patterns. We concluded that while a significant number of these patterns is still consistent with all the observations at 68.27\% C.L., there are several textures that can be excluded or are marginally allowed. We have also considered predictive neutrino zero textures with the assumption that the charged lepton mass matrix has the well-known NNI form. In the latter case, requiring non-Hermiticity of the charged lepton mass matrix is a necessary condition to obtain viable neutrino patterns. Predictive textures were found with a maximum number of three zeros, for both Majorana and Dirac neutrinos.  

It is well known that texture-zero models have in general a weak predictive power. We have not addressed here the question of the predictability of a given texture. This issue is beyond the scope of the present work. The reader is referred, e.g., to Ref.~\cite{Ludl:2014axa}, in which the authors attempted to identify predictive classes of texture zeros by defining numerical measures of predictability. For instance, maximally restrictive Majorana textures can predict, in most cases, the effective neutrino mass parameter $m_{\beta\beta}=|\sum_i \mathbf{U}_{ei}^2\,m_i|$, relevant in neutrinoless double beta decays.

From our study of different lepton mass matrix textures, it becomes clear that present neutrino oscillation data does not give an explicit hint on which category of textures is more appropriate to describe the observations. The precise measurements of neutrino oscillation parameters in upcoming experiments (including the determination of the absolute neutrino mass scale and the Dirac CP phase, and the improvement of the bounds on the sum of neutrino masses and the effective mass in $0\nu\beta\beta$ decays) are expected to shed some light on the flavor structure of the neutrino sector. This in turn would allow us to determine, among the plethora of texture-zero patterns, what are the most predictive textures capable of explaining the experimental data, as well as those that are disfavored or excluded at a high confidence level.

\section*{Acknowledgements}
We are grateful to M. Nebot and S. Palomares-Ruiz for useful discussions and comments. The work of D.E.C.~was supported by \emph{Associa\c c\~ao do Instituto Superior T\'ecnico para a Investiga\c c\~ao e Desenvolvimento} (IST-ID) and \textit{Funda\c{c}\~{a}o para a Ci\^{e}ncia e a Tecnologia} (FCT). D.E.C. and R.G.F. acknowledge support from FCT through the projects CERN/FP/123580/2011, PTDC/FIS-NUC/0548/2012 and  UID/FIS/00777/2013, and thank CERN Theoretical Physics Unit for hospitality and financial support.

\appendix

\section{Neutrino observables for the FGM two-zero textures}
\label{appendix1}

\begin{figure*}[h]
\centering
\includegraphics[width=0.57\linewidth]{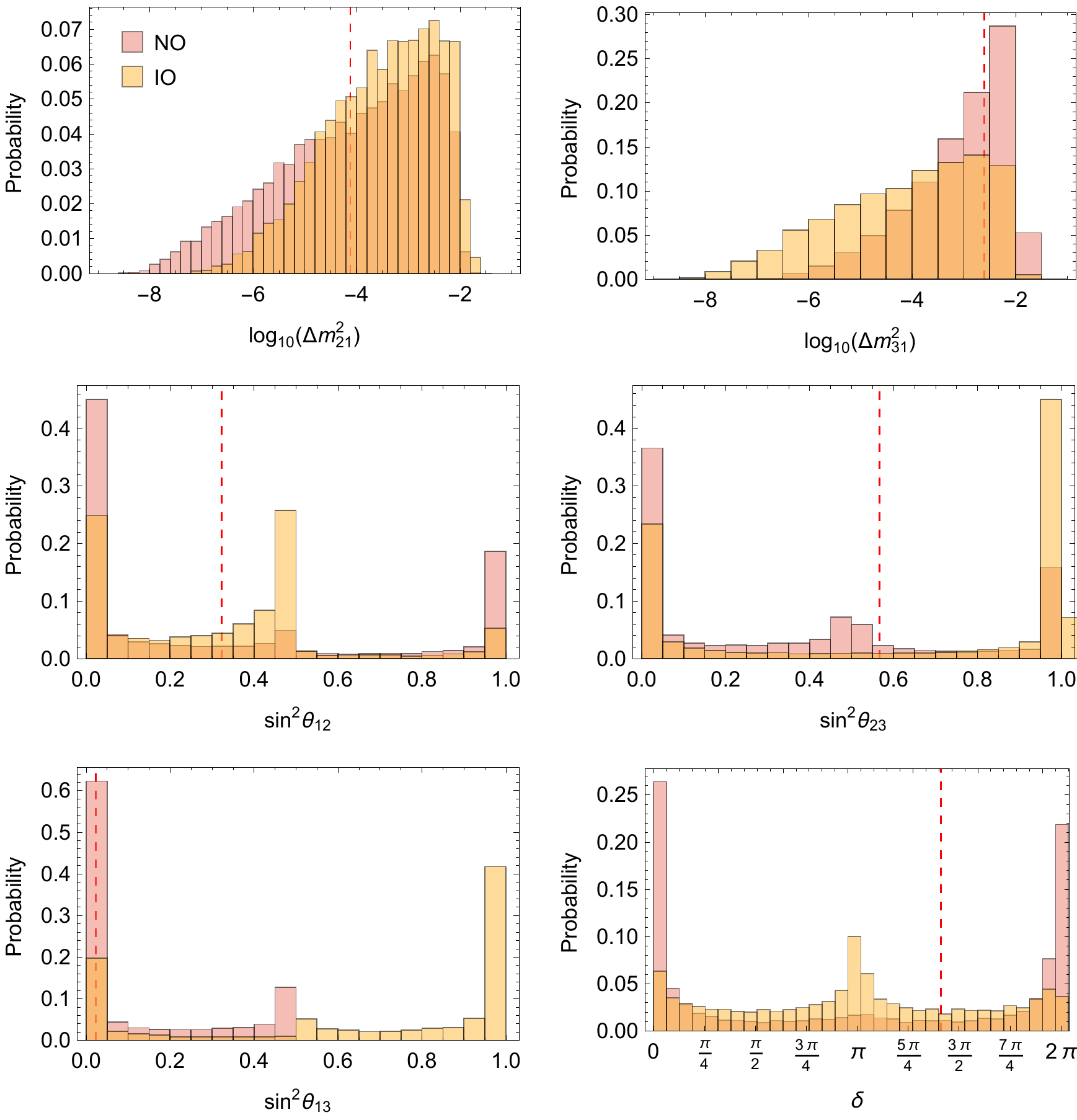}
\caption{The probability distribution of neutrino observables for pattern $\mathbf{A}_{1}$ in the case of Majorana neutrinos. The vertical dashed line denotes the best-fit value of the observable in the case of a normal ordering of the neutrino mass spectrum.}
\label{fig:fig-A1}
\end{figure*}

\begin{figure*}
	\centering
	\includegraphics[width=0.57\linewidth]{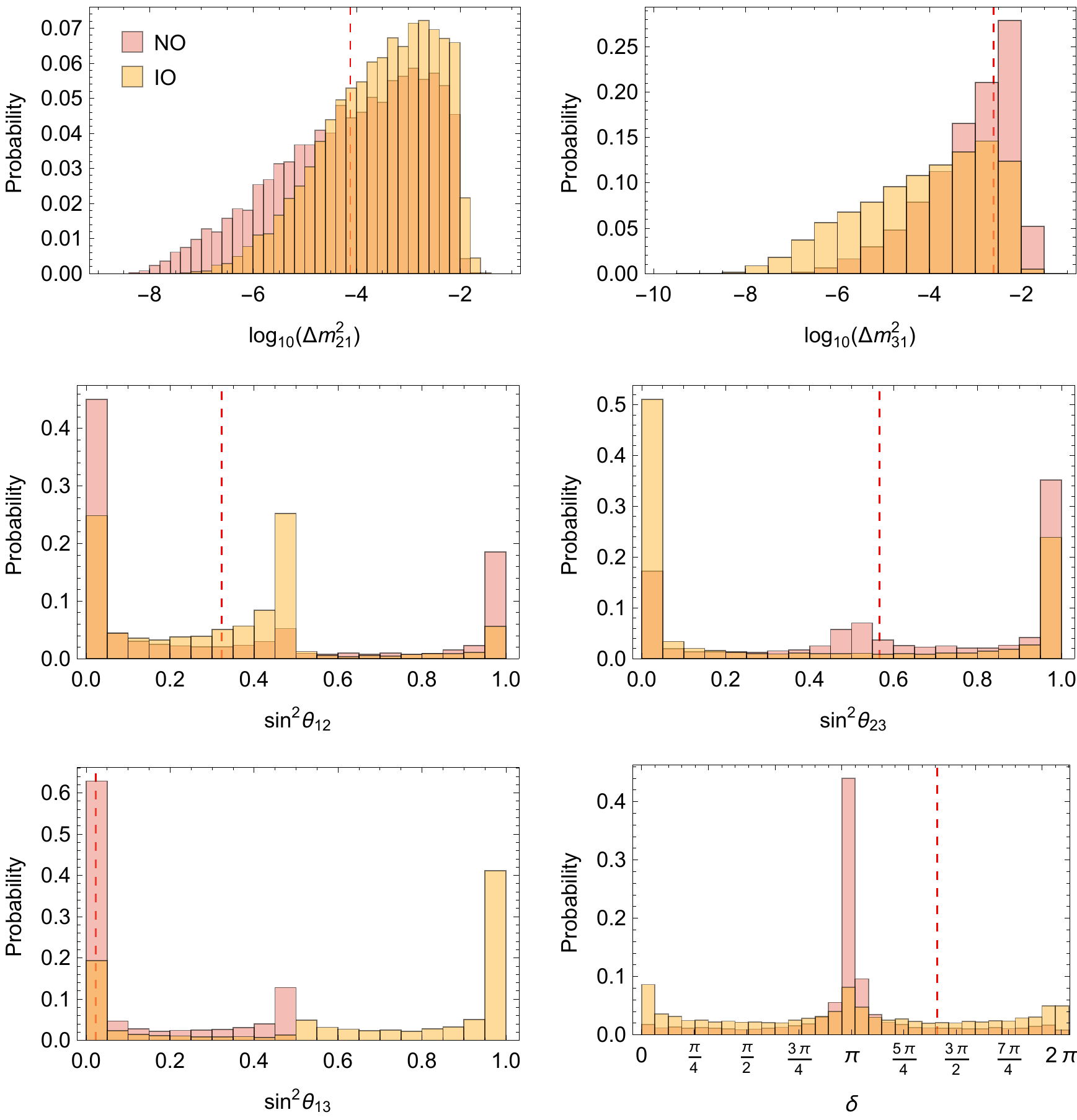}
	\caption{The probability distribution of neutrino observables for pattern $\mathbf{A}_{2}$ in the case of Majorana neutrinos.}
	\label{fig:fig-A2}
\end{figure*}

\begin{figure*}
\centering
\includegraphics[width=0.57\linewidth]{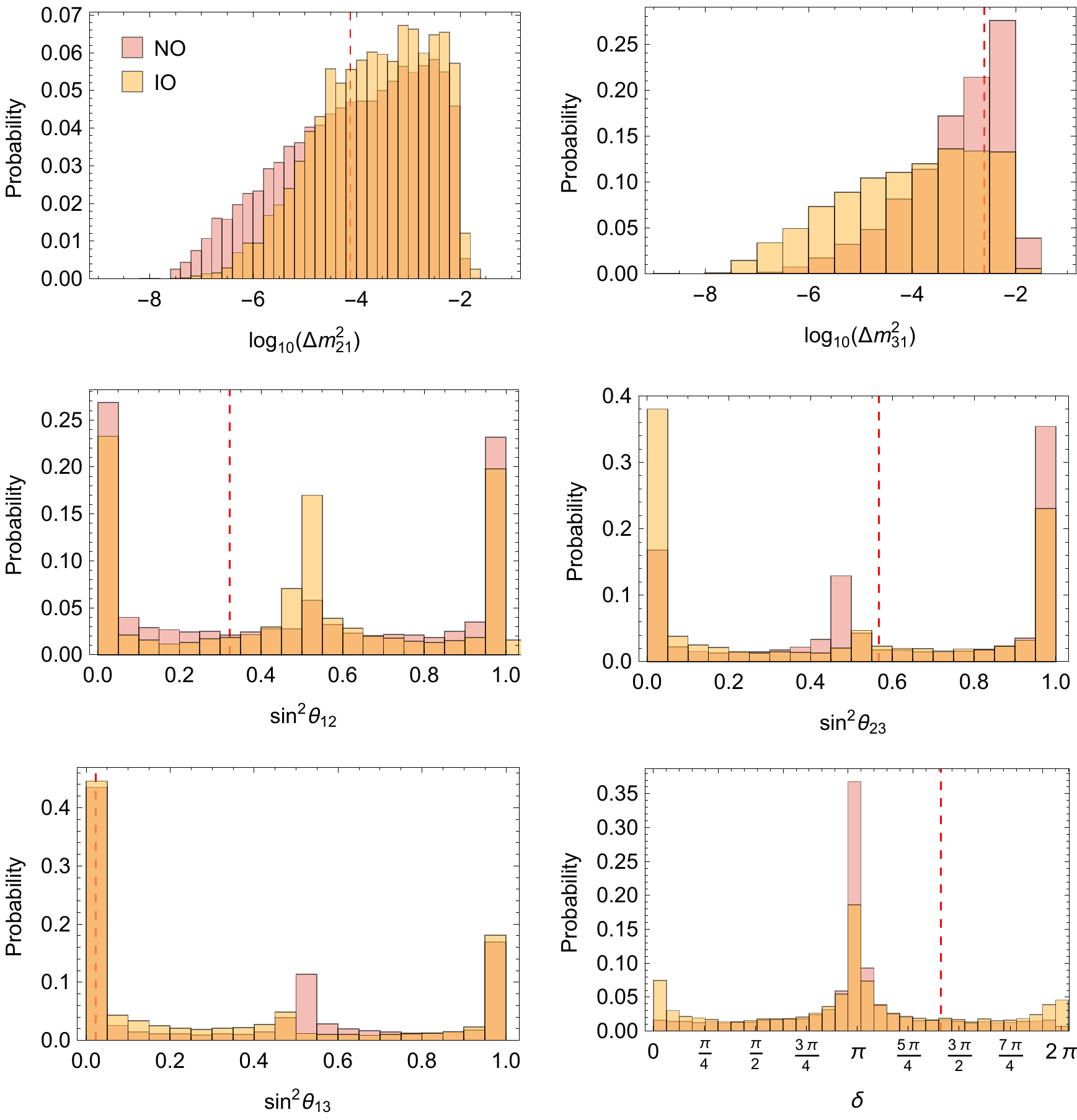}
\caption{The probability distribution of neutrino observables for pattern $\mathbf{B}_{1}$ in the case of Majorana neutrinos.}
\label{fig:fig-B1}
\end{figure*}

\begin{figure*}
	\centering
	\includegraphics[width=0.57\linewidth]{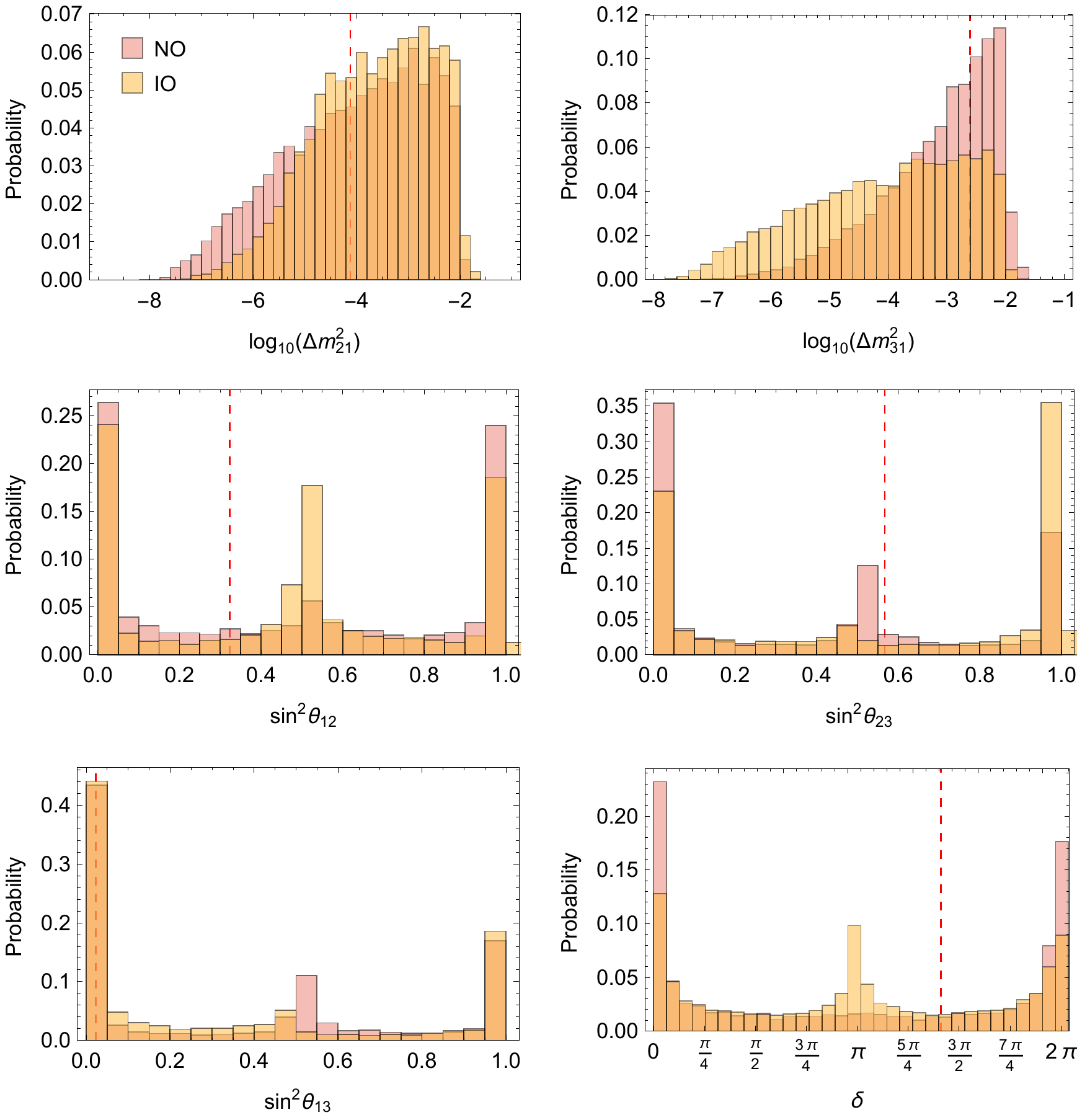}
	\caption{The probability distribution of neutrino observables for pattern $\mathbf{B}_{2}$ in the case of Majorana neutrinos.}
	\label{fig:fig-B2}
\end{figure*}

\begin{figure*}
	\centering
	\includegraphics[width=0.57\linewidth]{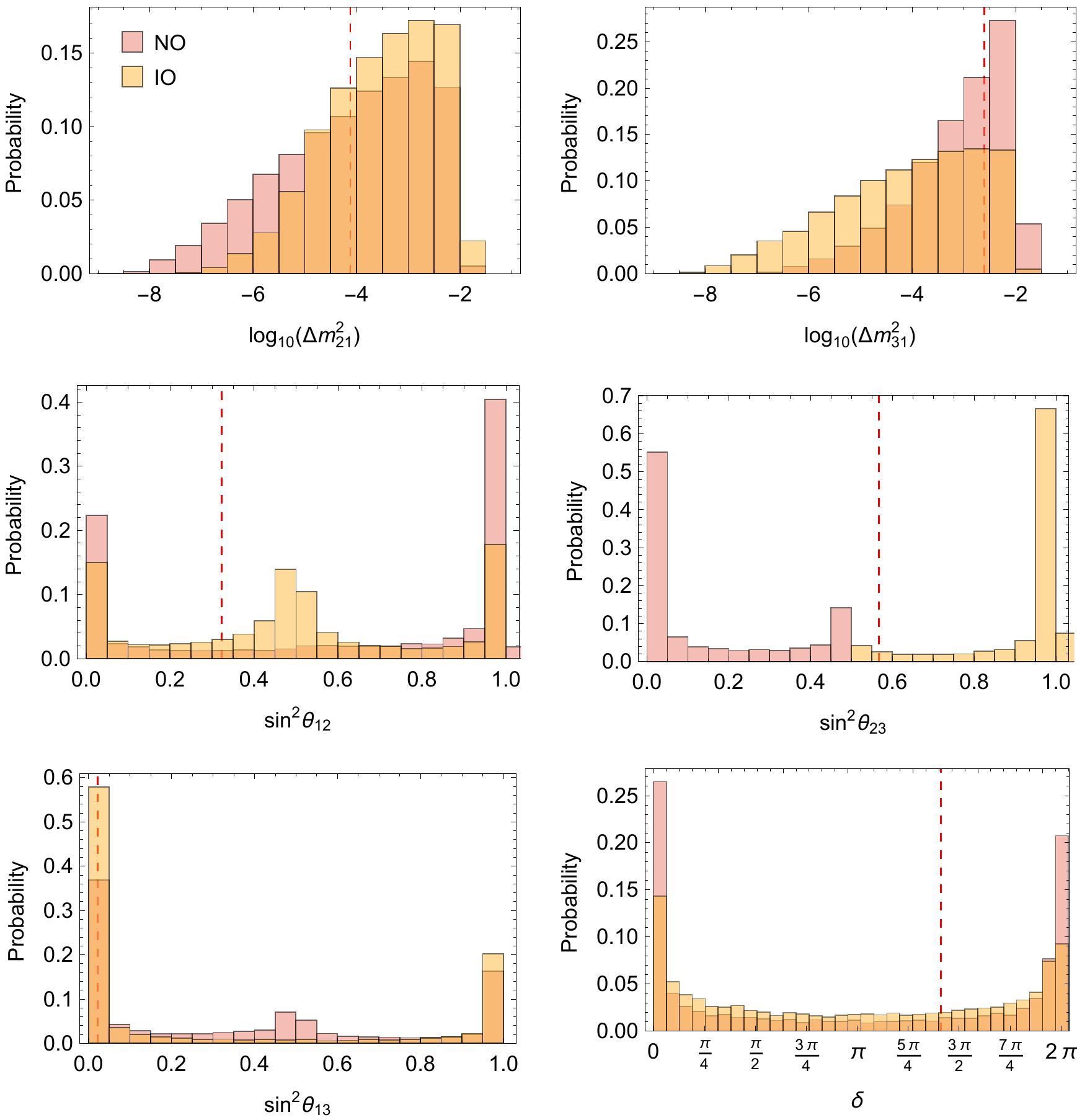}
	\caption{The probability distribution of neutrino observables for pattern $\mathbf{B}_{3}$ in the case of Majorana neutrinos.}
	\label{fig:fig-B3}
\end{figure*}

\begin{figure*}
	\centering
	\includegraphics[width=0.57\linewidth]{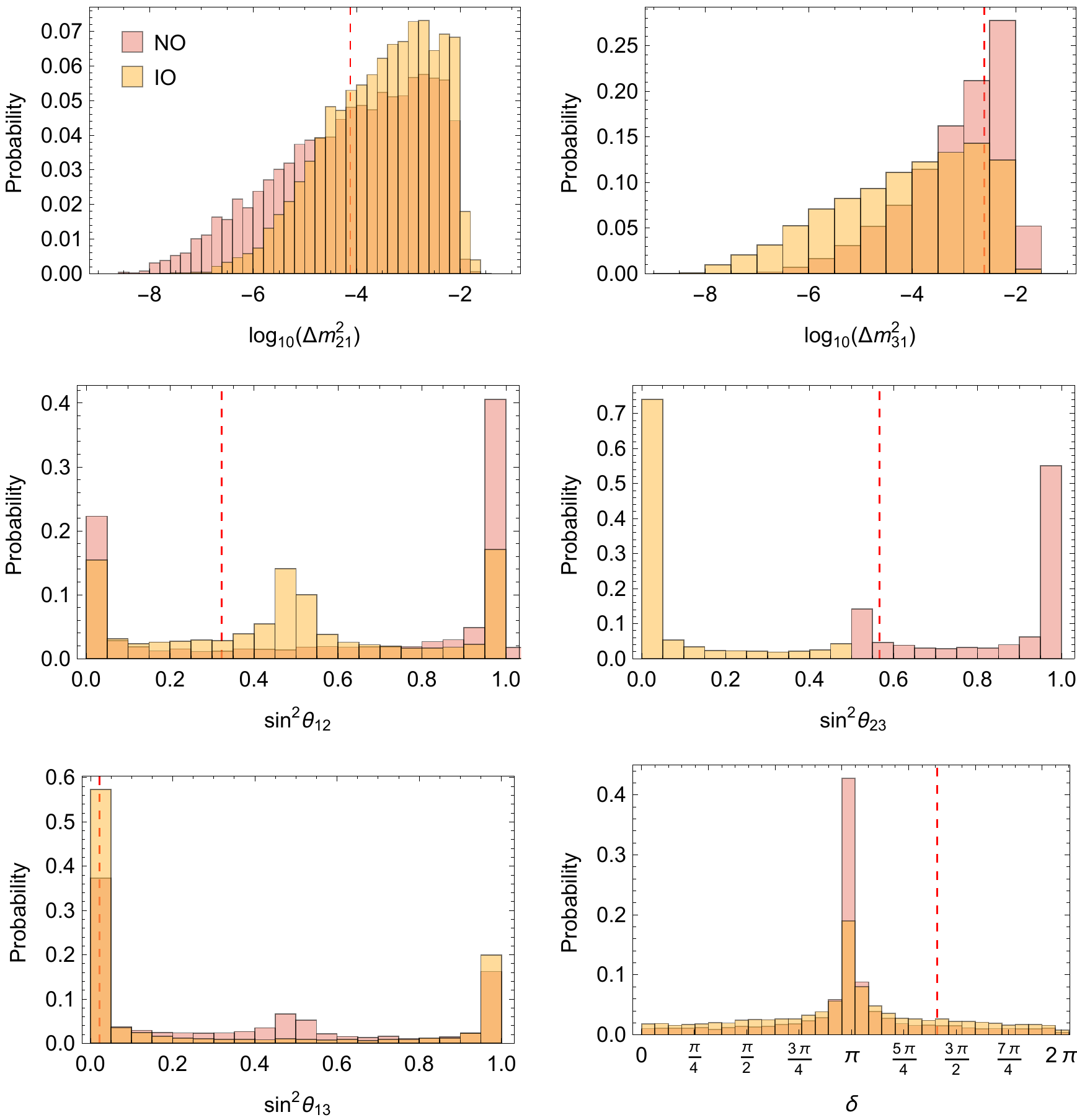}
	\caption{The probability distribution of neutrino observables for pattern $\mathbf{B}_{4}$ in the case of Majorana neutrinos.}
	\label{fig:fig-B4}
\end{figure*}

\begin{figure*}
	\centering
	\includegraphics[width=0.57\linewidth]{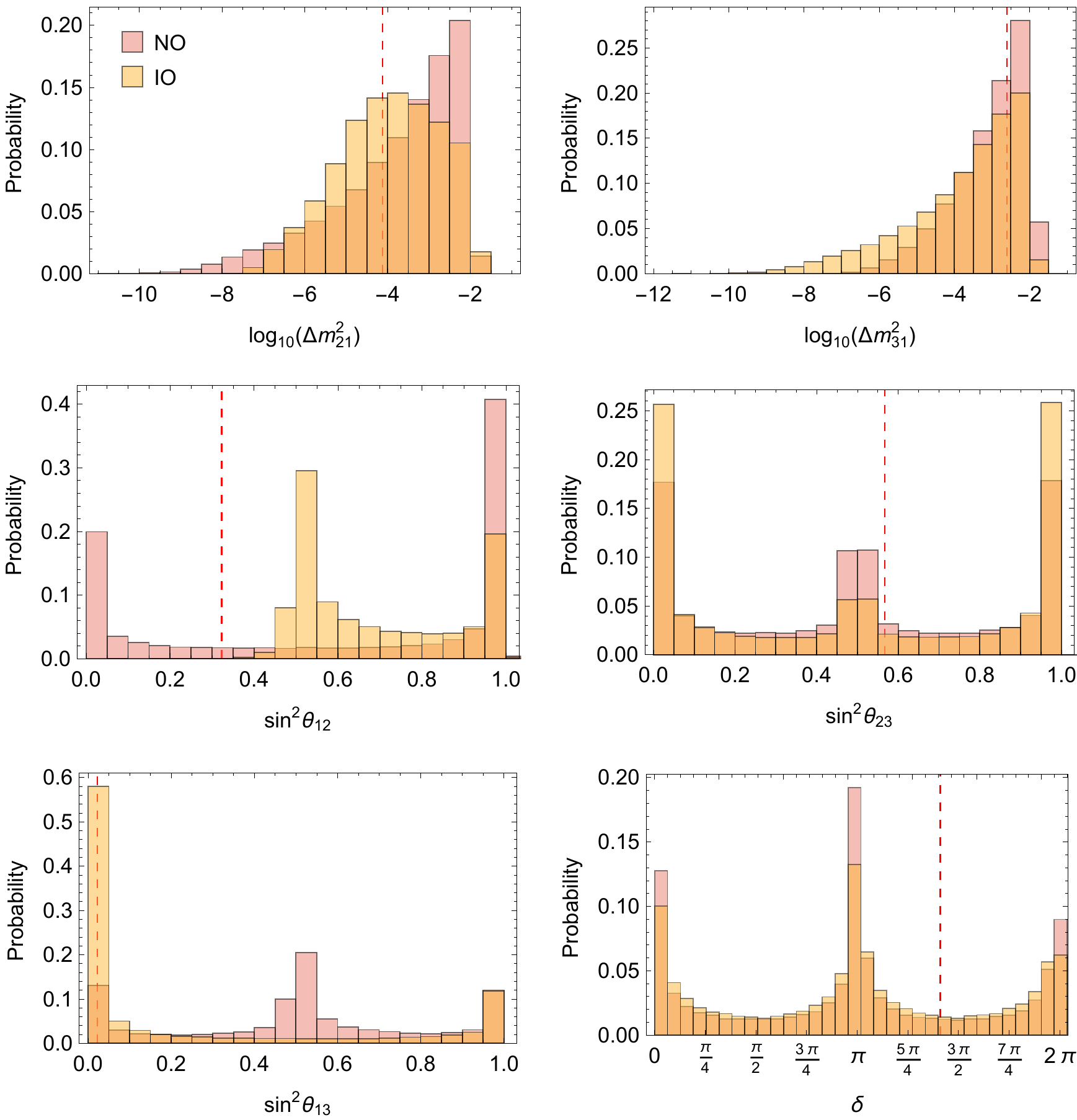}
	\caption{The probability distribution of neutrino observables for pattern $\mathbf{C}$ in the case of Majorana neutrinos.}
	\label{fig:fig-C}
\end{figure*}

For completeness, in this appendix we present the probability distribution of the neutrino observables for the viable FGM patterns $\mathbf{A}_{1,2}$, $\mathbf{B}_{1,2,3,4}$ and $\mathbf{C}$ for NO and IO mass spectra. The results are presented in Figs.~\ref{fig:fig-A1}-\ref{fig:fig-C} and have been obtained using $10^4$ random input neutrino mass matrices. In all figures, the vertical red dashed line denotes the best-fit value of the corresponding observable, taken from Table~\ref{tab:nudata} for a normal ordering of the neutrino mass spectrum. The histograms reflect the frequency of the values of the six neutrino observables included in the $\chi^2$-analysis, namely, the mass-squared differences $\Delta m_{21}^2,\, \Delta m_{31}^2$, the mixing angles $\mathrm{sin}^2\theta_{12},\, \mathrm{sin}^2\theta_{23},\,
\mathrm{sin}^2\theta_{13}$, and the Dirac CP phase $\delta$.

\vspace*{1mm}
\section{Neutrino observables for three-zero parallel textures}

\begin{figure*}
\centering
\includegraphics[width=0.57\linewidth]{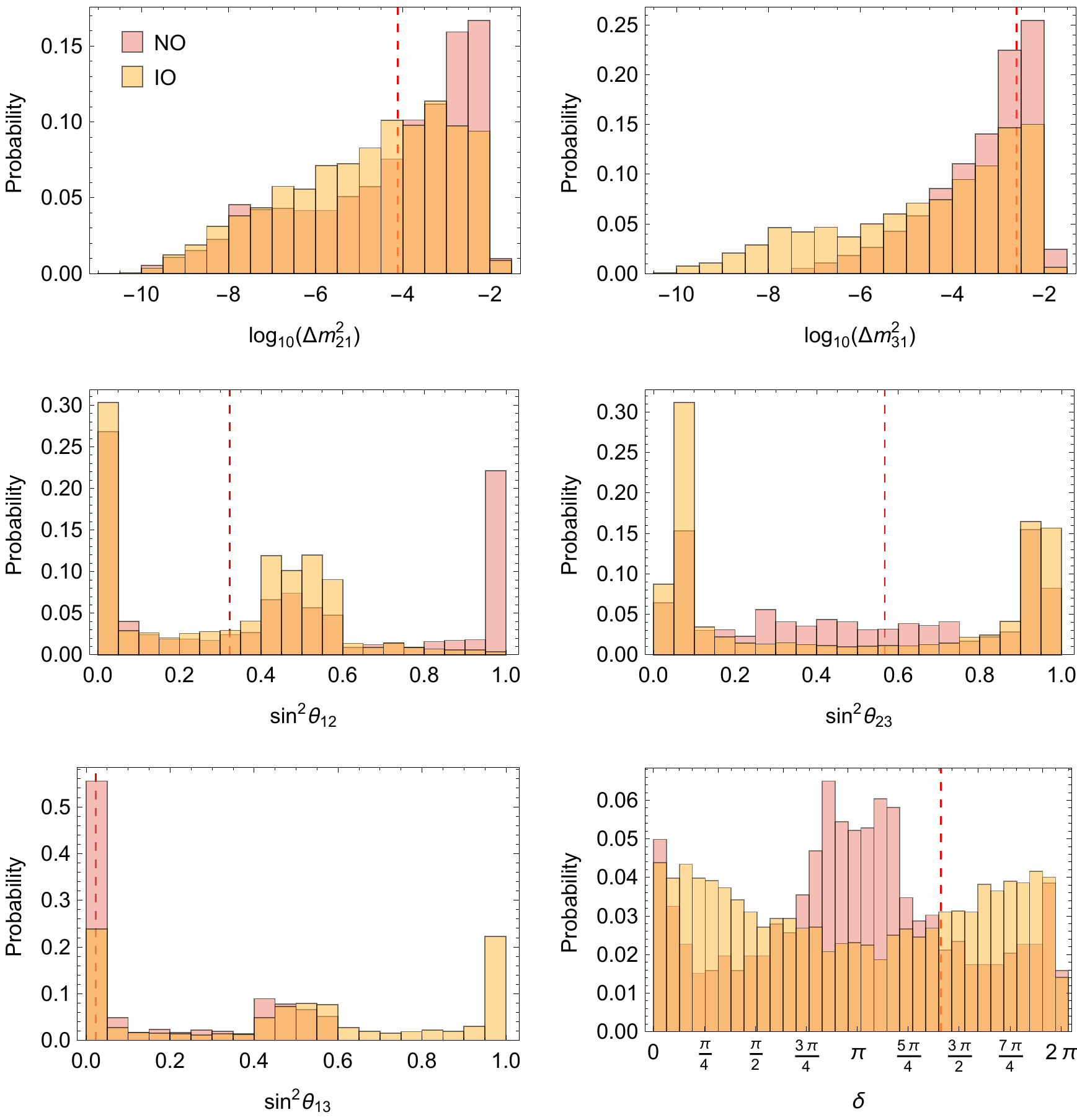}
\caption{The probability distribution of neutrino observables for the three-zero parallel patterns given in Eq.~\eqref{3zerotextures}, in the case of Majorana neutrinos. Similar results are obtained for Dirac neutrinos.}
\label{fig:fig-3zeros-Majorana}
\end{figure*}

Here we present the probability distribution of neutrino observables for the three-zero parallel patterns given in Eq.~\eqref{3zerotextures}, in the case of Majorana neutrinos. As in the case of the FGM textures, $10^4$ random input mass matrices for charged leptons and neutrinos were generated. The results are presented in Fig.~\ref{fig:fig-3zeros-Majorana}. Similar distributions are obtained if neutrinos are Dirac particles.

\label{appendix2}

\end{document}